# Imaging Flux Vortices in Type II Superconductors with a Commercial Transmission Electron Microscope

## Authors

J.C. Loudon and P.A. Midgley

## Affiliation

Department of Materials Science and Metallurgy
University of Cambridge
Pembroke Street
Cambridge CB2 3QZ, UK

## Abstract

Flux vortices in superconductors can be imaged using transmission electron microscopy because the electron beam is deflected by the magnetic flux associated with the vortices. This technique has a better spatial and temporal resolution than many other imaging techniques and is sensitive to the magnetic flux density within each vortex, not simply the fields at the sample surface. Despite these advantages, only two groups have successfully employed the technique using specially adapted instruments. Here we demonstrate that vortices can be imaged with a modern, commercial transmission electron microscope operating at 300 kV equipped with a field emission gun, Lorentz lens and a liquid helium cooled sample holder. We introduce superconductivity for non-specialists and discuss techniques for simulating and optimising images of flux vortices. Sample preparation is discussed in detail as the main difficulty with the technique is the requirement for samples with very large (>10μm), flat areas so that the image is not dominated by diffraction contrast. We have imaged vortices in superconducting $Bi_2Sr_2CaCu_2O_{8-\delta}$ and use correlation functions to investigate the ordered arrangements they adopt as a function of applied magnetic field.

## 1. Introduction

This paper discusses techniques for imaging flux vortices in superconductors using transmission electron microscopy and we demonstrate that vortices can be imaged using a modern, commercially available microscope. It also introduces the basics of superconductivity for those interested in the techniques but unfamiliar with the subject.

The defining features of a superconductor are that it has zero electrical resistivity and it expels magnetic flux from its interior (the Meissner effect) [1]. This behaviour is explained by the Bardeen-Cooper-Schrieffer (BCS) model of 1957 [2] which demonstrates that a coupling between electrons and vibrations of the crystal lattice can lead to an attractive interaction between the conduction electrons so that they form bound pairs called Cooper pairs. The size of each Cooper pair is called the coherence length, $\xi$, and it can vary between 1 and 1000 nm depending on the superconductor [3]. The Cooper pairs form a 'condensate' containing a macroscopic number of pairs, all in the same quantum state. Although electrons can be thermally scattered into and out of the condensate, a macroscopic number remain condensed in a constant quantum state so that energy cannot be dissipated and the superconducting electrical currents do not die away. As the details of BCS theory are mathematically challenging, it is usual to introduce superconductivity using the phenomenological models which preceded it [4].

The phenomenological models begin with a consideration of the Drude model of electrical conductivity [4] which uses Newton's second law to give the velocity (**v**) of an electron in a solid under the influence of an electric field (**E**).

$$m\frac{d\mathbf{v}}{dt} = -e\mathbf{E} + \mathbf{F}_{\text{scatter}}(t) \tag{1}$$

where $m$ = electron mass, $t$ = time, $e$ = electron charge and **E** = electric field,

In a normal material, electrons are scattered by defects and phonons and this is accounted for by the scattering force, $\mathbf{F}_{\text{scatter}}(t)$. After a few collisions, the force from the electric field is balanced by the scattering force so that the acceleration averages to zero and the electrons move at a constant speed when a constant electric field is applied.

In a superconductor, however, there is no scattering force. The phenomenological models do not attempt to explain this but take it as an axiom in order to investigate the consequences. The first consequence is that under a constant electric field, the electrons will accelerate without limit as expressed by the first London equation [5]

$$\frac{d\mathbf{J}}{dt} = \left(\frac{n_s e^2}{m}\right)\mathbf{E} \tag{2}$$

This is simply equation (1) without the scattering force expressed in terms of the current density $\mathbf{J} = -n_s e\mathbf{v}$ where the number density of electrons involved in superconductivity is denoted $n_s$.

A reader wondering how a controlled current can be applied to a superconductor if the electrons accelerate without limit in an electric field should note that this behaviour is the same as an ideal, resistanceless wire. To apply a controlled current, it is necessary to use an impedance in series otherwise the current will become very high and any resistive components such as the connections to the power supply will become very hot. The advantage of using superconducting wires is that they can carry greater currents than conventional wires before heating becomes a problem.

The flux density **B** within a superconductor can be investigated by taking the curl of equation (2) and using Maxwell's third equation, $\nabla \times \mathbf{E} = -\frac{\partial \mathbf{B}}{\partial t}$, to give

$$\frac{d(\nabla \times \mathbf{J})}{dt} = -\left(\frac{n_s e^2}{m}\right)\frac{\partial \mathbf{B}}{\partial t} \tag{3}$$

or, integrating both sides with respect to time,

$$\nabla \times \mathbf{J} + \left(\frac{n_s e^2}{m}\right)\mathbf{B} = \mathbf{C} \tag{4}$$

where **C** is a constant vector.

It should be noted that there is an error in this last step usually ignored in textbooks. There are two types of time derivative here: $d/dt$ is the convective derivative which expresses the rate of change in a frame moving with a particular electron whereas the stationary derivative $\partial/\partial t$ is the rate of change at a fixed point in space. For an arbitrary vector **V**, the two are related via $d\mathbf{V}/dt = \partial \mathbf{V}/\partial t + (\mathbf{v}.\nabla)\mathbf{V}$. It turns out that for most geometries, the $(\mathbf{v}.\nabla)\mathbf{V}$ term is negligible [6] and equation (4) is a good approximation.

To obtain an equation involving only **B**, Maxwell's fourth equation is used

$$\nabla \times \mathbf{B} = \mu_0 \mathbf{J} + \mu_0 \varepsilon_0 \frac{\partial \mathbf{E}}{\partial t} \tag{5}$$

and we can investigate steady state solutions by setting the time derivatives equal to zero, taking the curl of (5) and substituting the result into (4) to give

$$\nabla \times (\nabla \times \mathbf{B}) + \left(\frac{n_s e^2 \mu_0}{m}\right)\mathbf{B} = \mathbf{C}. \tag{6}$$

Recalling the vector identity $\nabla \times (\nabla \times \mathbf{B}) = \nabla(\nabla.\mathbf{B}) - \nabla^2 \mathbf{B}$ and that $\nabla.\mathbf{B} = 0$ we obtain

$$-\nabla^2 \mathbf{B} + \left(\frac{n_s e^2 \mu_0}{m}\right)\mathbf{B} = \mathbf{C} \tag{7}$$

We can gain an insight into the behaviour of the B-field in a superconductor by considering the problem of a semi-infinite slab of superconductor occupying the space $x > 0$ with a uniform B-field applied in the $x$ direction. Since the B field is now only ever in the $x$ direction, the B-field within the superconductor is the solution to the 1-dimensional equation

$$-\frac{d^2B}{dx^2} + \left(\frac{n_s e^2 \mu_0}{m}\right) B = C \qquad (8)$$

which is an ordinary second order differential equation. The solution to this is a complementary function which solves the case where the constant is zero added to a particular integral which is any solution to the complete equation. It can be seen that a constant particular integral, $B_0$, will solve the equation and that the complementary function has an exponential form. Applying the boundary condition that the B-field cannot diverge with increasing $x$ gives the solution

$$B(x) = B_0 + B_1 \exp(-x/\Lambda) \qquad (9)$$

where $B_1$ is constant and

$$\Lambda = \sqrt{m/n_s e^2 \mu_0} . \qquad (10)$$

Whilst this expression is correct based on the assumptions we have made so far, it was demonstrated by the experiments of Meissner and Ochsenfeld [1] that the B-field is always zero in the interior of a superconductor. This is not explained by the phenomenological models and is regarded as a further axiom. To make the model consistent with the experiment, it is necessary that $B_0$ be zero. Thus a B-field applied to a superconductor will tend to decay exponentially to zero from its surface over a characteristic length $\Lambda$, called the penetration depth. Values of $\Lambda$ range from 30–300 nm depending on the superconductor [3]. Tracing back through the preceding discussion, the condition that there must be no B-field in the interior of the superconductor implies that the constant in equation (4) must be zero. This gives the second London equation:

$$\nabla \times \mathbf{J} + \mu_0 \mathbf{B}/\Lambda^2 = 0 \qquad (11)$$

The two London equations together express the macroscopic electrodynamics of superconductors.

There is an upper limit to the magnetic flux that a superconductor can expel and when a magnetic field is applied to a superconductor which is larger than the so-called critical field, $H_C$, the superconducting state is destroyed, magnetic flux enters and the sample becomes normal (i.e. non-superconducting). Such materials are termed type I superconductors. Typical values of $H_C$ at 0 K range from 0.1–80 mT in elemental superconductors [7]. However there are also superconductors, termed type II superconductors, where the superconducting state is not entirely destroyed when magnetic flux enters. Instead, superconducting and normal regions coexist. Type II superconductors are characterised by two critical fields: the lower critical field $H_{C1}$ and the upper critical field $H_{C2}$. When a field is applied below $H_{C1}$, no flux enters the interior of the sample; when the field is between $H_{C1}$ and $H_{C2}$, superconducting and normal regions coexist and above $H_{C2}$, the superconducting state is completely destroyed. It can be shown that in type I superconductors $\Lambda/\xi < 1/\sqrt{2}$ and in type II $\Lambda/\xi > 1/\sqrt{2}$ [3].

In niobium, one of the few elemental type II superconductors, $H_{C1}$ = 139 mT and $H_{C2}$ = 268 mT at 4.2 K [7]. In compounds, the upper critical field can be much larger such as Nb$_3$Sn which has $H_{C2}$ = 25 T at 4.2 K [7]. In high temperature superconductors (so called because of their unusually high superconducting transition temperature), the upper critical field can be larger still. In Bi$_2$Sr$_2$CaCu$_2$O$_{8-\delta}$ (BSCCO), the superconductor used in this investigation, $H_{C2}$ = 280 T at 0 K for a field applied parallel to its superconducting planes and 32 T for a perpendicular field [8]. It should be noted that the critical fields refer to samples in which there are no pinning forces as pinning effects can prevent flux entering the sample at fields considerably higher than the lower critical field.

The reason for the two distinct types of superconductor is the energy associated with forming an interface between normal and superconducting regions. In a type I superconductor, the energy is positive and so the interfacial area is as small as possible. In a type II superconductor, however, the interfacial energy becomes negative above $H_{C1}$ and, as Abrikosov [9] first showed, flux penetrates type II superconductors along many cylindrical channels, called flux vortices, each containing the smallest amount of magnetic flux consistent with the quantum mechanical description of superconductivity. As a result of making the quantum mechanical wavefunction associated with the superconducting state single valued, it can be shown that the total magnetic flux associated with each vortex is $\Phi_0 = h/2e = 2.07 \times 10^{-15}$ Tm$^2$ [3,9]. This quantity is called the flux quantum and the $2e$ is a consequence of the charge on the Cooper pairs which make up the superconducting state. These flux vortices are fundamental particles in the sense that each vortex always has the same total magnetic flux associated with it and it is not possible to subdivide this. To emphasise their quantum nature, they are frequently called fluxons.

Figure 1 shows a schematic diagram of fluxons in a superconductor generated by a uniform applied B-field. Each vortex consists of a cylindrical core with a radius characterised by the coherence length, $\xi$, where the density of Cooper pairs is greatly reduced, falling to zero at the centre. (In fact, the line on which the Cooper pair density falls to zero defines the centre of the vortex core and this is true for all flux vortices including the Josephson vortices discussed in section 4.) Electrical superconducting currents (supercurrents), **J**, circulate around the core and have a characteristic radius given by the penetration depth $\Lambda$. These currents concentrate the B-field within a radius $\Lambda$ of the core. Flux vortices are mutually repulsive due to the Lorentz force between them and so in an infinitely large, defect-free, isotropic sample, they adopt a hexagonal lattice known as an Abrikosov lattice as this has the highest packing fraction [9].

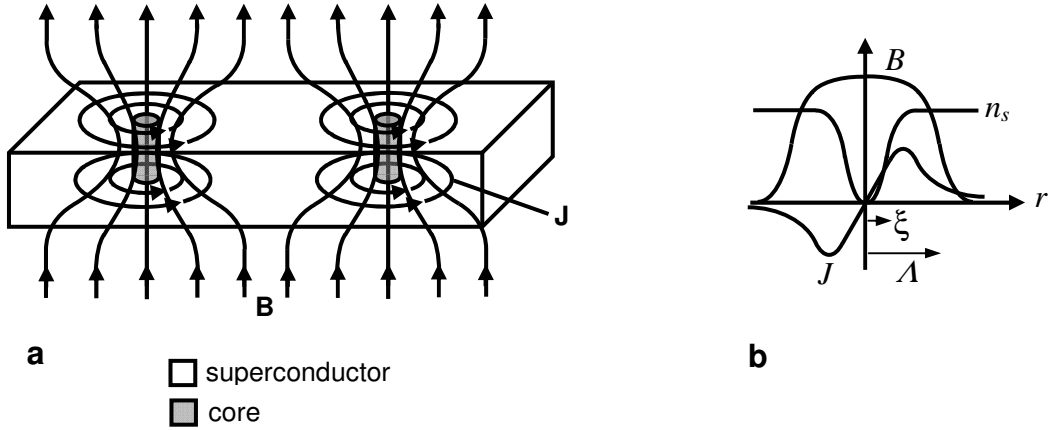

**Figure 1:** (a) Schematic diagram showing magnetic flux penetrating a superconductor via flux vortices. (b) Schematic cross section showing the B-field, number density of electrons contributing to superconductivity, $n_s$ and supercurrent $J$ as a function of distance $r$ from the centre of a vortex.

In the absence of pinning forces, the density of vortices in the specimen is such that the flux density averaged over a large area in the superconductor is the same as the applied B-field. In a square lattice this would mean that the vortex spacing would be $s = \sqrt{\Phi_0 / B}$. For a hexagonal (Abrikosov) lattice of vortices, the spacing is

$$s = \sqrt{\frac{\sqrt{3}}{2}\frac{\Phi_0}{B}} \qquad (12)$$

The behaviour of fluxons under an external influence is of great importance for superconductor-based technology such as electromagnets for magnetic resonance imaging made with superconducting wires. If a sufficiently large electrical current is passed through a type II superconductor, flux vortices are generated by the magnetic field associated with the current as well as any external fields which may be present. There is then a Lorentz force (called the Magnus force) on the vortices due to the current

$$\mathbf{f} = \mathbf{J} \times \mathbf{\Phi_0} \qquad (13)$$

where $\mathbf{f}$ is the force per unit length of the fluxon, $\mathbf{J}$ the current density and $\mathbf{\Phi_0}$ is the magnetic flux associated with the fluxon pointing in the direction of the magnetic field. If the fluxons move in response to this force, work is done and so energy is dissipated and the sample behaves as though it had a finite electrical resistance even though it is superconducting.

Vortices can only move freely in an ideal sample. Defects which lead to sudden changes in the superconducting state such as grain boundaries and point defects can pin the vortices. It should be noted that vortices can only be created or destroyed at the sample surfaces and when the B-field is turned off, they tend to migrate to the sample surface where they are destroyed. Pinning forces can cause vortices to remain in the sample for significant lengths of time even when the B-field is turned off, leading to magnetic hysteresis. Pinning forces inhibit the movement of vortices and this reduces energy losses when a superconductor is used to carry an electrical current. Thus, there is a great deal of interest in the movement and pinning of flux vortices and electron microscopy has the potential to study the dynamics of vortices and clarify the nature of the sites which pin vortices best.

Figure 2 shows the basic experimental arrangement used here for imaging vortices with a transmission electron microscope. The method relies on the fact that flux vortices tend to penetrate thin film samples approximately normal to the thin film surface of the superconductor as discussed in section 4. Thus, when the sample is tilted with respect to the electron beam as shown in Figure 2, there is a component of the B-field normal to the electron beam so that the electrons are deflected and the fluxons are visible as black-white features in an out-of-focus (Fresnel) image. We now summarise the historical development of this technique.

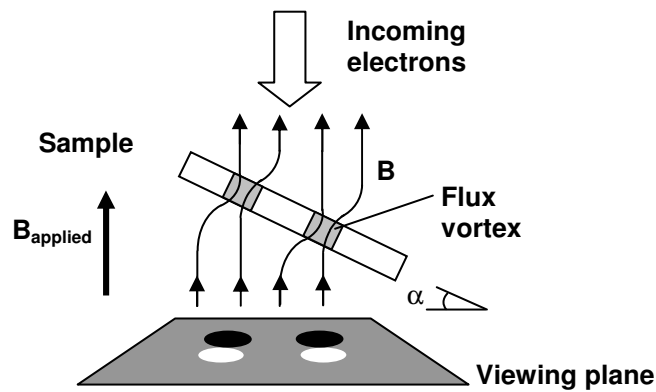

**Figure 2:** Experimental arrangement for the observation of flux vortices. The electrons are deflected by the component of the field normal to the beam producing black-white features in an out-of-focus image.

## 2. History of Fluxon Imaging using Transmission Electron Microscopy

Here we summarise the history of vortex imaging by transmission electron microscopy. It was first suggested that vortices could be imaged with an electron microscope by Hirsch, Howie and Jakubovics in 1965 [10]. Following this, there were several theoretical calculations on whether it might be possible to image vortices. In 1966, Yoshioka [11] calculated the appearance of Fresnel and Foucault images of vortices lying normal to the electron beam and a similar calculation was made by Wohlleben in 1967 [12]. In 1968, Colliex *et al.* [13] considered the case of fluxons with their axes parallel to the electron beam and concluded that the resulting phase shifts would be undetectably small. Capiluppi *et al.* [14] (1972) calculated the appearance of Fresnel images for vortices normal to the electron beam and concluded that in this case, they should be experimentally detectable. As expected then, flux vortices are most visible when their axes are normal to the electron beam as the Lorentz force on the electrons is largest when the B-field is normal to the electron beam.

The problem is that, as mentioned in the introduction, vortices tend to take the shortest route through a specimen and for an electron microscopy specimen, their axes will point within a few degrees of the normal to the thin surface (see section 4) even if the magnetic field is applied parallel to the thin surface (see Figure 3(a)). In order to orient the vortices normal to the electron beam the specimen must be positioned so the electron beam runs normal to the thin direction as shown in Figure 3(b): exactly the opposite configuration to that of almost all other microscopy experiments. This has the obvious disadvantage that the specimen is no longer electron transparent and so only the fields in the vacuum can be observed.

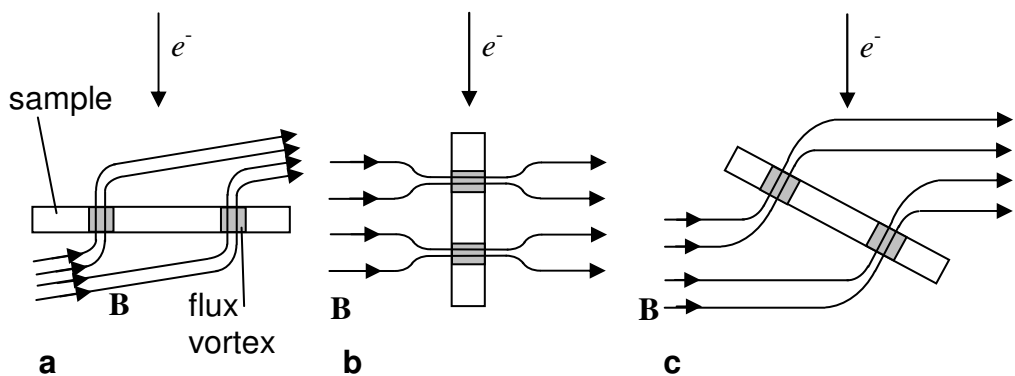

**Figure 3:** Configurations for experimental observation of fluxons using electron microscopy. (a) A field which deviates only slightly from the plane of the thin sample will still induce approximately vertical fluxons and the vortices are almost invisible because the flux density in their core is parallel to the electron beam. (b) The flux density can be made perpendicular to the electron beam by aligning the sample as shown. However, the sample is now thick in the electron beam direction and only stray fields can be observed. (c) The best configuration is to have sample tilted with respect to the beam.

One of the main obstacles to imaging vortices at this time was the need for coherent illumination as the deflection angles induced by flux vortices are of the order of micro-radians. Spatial coherence can always be achieved by reducing the size of the condenser aperture but on microscopes with $LaB_6$ and tungsten filaments, the beam would then be too dim to detect vortices. With the development of the field emission gun, the gun brightness was increased from $10^5$-$10^6$ A cm$^{-2}$ sr$^{-1}$ to ~$10^9$ A cm$^{-2}$ sr$^{-1}$ and it became feasible to image vortices.

Flux vortices were first observed in 1989 in superconducting Pb by Matsuda *et al.* [15] working in the Tonomura group at the Hitachi Advanced Research Laboratory in Japan using electron holography with the specimen in the vertical configuration shown in Figure 3(b). The presence of vortices in the sample was inferred from the observation of stray magnetic fields in the vacuum. It should be noted that Pb is actually a type I superconductor but it can be demonstrated that when a field is applied normal to the thin surface of a superconductor, magnetic flux penetrates as filaments of quantised flux although some of the filaments contain several quanta, not simply one as would be the case in a type II superconductor [16].

The restriction of being unable to see the vortices in the superconductor and seeing only their stray fields can be circumvented by using a sample tilted not to 90° but to an intermediate angle (Figure 3(c)) and shortly after in 1992, this method was employed by Harada *et al.* [17] (Tonomura group) to take Fresnel images of vortices in the type II superconductor, Nb.

The microscope used in this experiment was a Hitachi H-9000 microscope operating at 300 kV which was modified from the standard design so that it had a field emission gun [18]. The sample was cooled to 4.5 K using an in-column liquid helium cooling stage [19] and tilted to 45°. The main objective lens of an electron microscope applies a field of ~2 T to a sample and so Fresnel images were taken by turning off the main objective lens and using the intermediate lens to focus. Images were recorded on photographic plates and in real time on a high gain TV camera. In this experiment and many of the others performed by Tonomura's group, the fluxons were generated by a horizontally applied magnetic field generated by small coils on either side of the holder. A horizontal field has the unwanted effect of deflecting the electron beam and there were similar coils below the specimen to compensate for the original deflection. It is not necessary to use a horizontal field because, as discussed earlier, vortices tend to align themselves approximately normal to the surface of the microscopy specimen. In our work (and also that of ref. 20), a vertical field is used which works equally well and does not deflect the electron beam. This method also has the advantage that the objective lens can be used to produce a vertical field and so a magnetising stage is not necessary. It should be emphasised, however, that these two geometries are distinct and that in situations where the vortex is does not take the shortest path through the material (described at the end of section 4), they will produce different results.

In 1998, the Horiuchi group at the National Institute for Research in Inorganic Materials in Japan imaged fluxons in Nb [21, 22]. The microscope used was a 300kV Hitachi HF-3000L with a field emission gun which was modified so that it had a Lorentz lens rather than the conventional objective. A Lorentz lens is designed so that it does not apply a field to the sample and although the lens aberrations are considerably worse than the objective lens (the coefficient of spherical aberration is ~8 m rather than ~1 mm), the resolving power is better than the intermediate lens. The specimen was cooled to 5 K using side-entry liquid helium stage and a horizontal magnetic field was applied by field coils built into the microscope column. Images were recorded on imaging plates instead of photographic film as these have the advantage that the intensity recorded depends linearly on the intensity of the electron beam and they are more sensitive than film so that exposure times can be reduced by a factor of ~6. To date, the Tonomura and Horiuchi groups have been the only ones to have successfully imaged flux vortices using this technique.

In 2000, the Tonomura group developed the first 1 MV electron microscope with a field emission gun which had a record gun brightness of $2\times10^{10}$ A cm$^{-2}$ sr$^{-1}$ [23, 24]. The higher voltage allowed thicker specimens to be observed which enhances the contrast seen from vortices. This and the other advantages of using a higher acceleration voltage are discussed in detail in section 5. The microscope has a liquid helium cooled in-column specimen stage equipped with coils to allow magnetic fields to be applied to the sample at any angle. This proved useful in the investigation of stacks of pancake vortices in YBCO (YBa$_2$Cu$_3$O$_{7-\delta}$) which apparently broke into several stacks as the angle at which the field was applied increased [25].

The Philips CM300 electron microscope used in this investigation uses a Schottky field emission gun operated at 300kV, for which the brightness is typically $10^9$ A cm$^{-2}$ sr$^{-1}$, and is equipped with a Lorentz lens. Images were recorded digitally with a CCD camera and, unlike previous fluxon images, are energy-filtered using a Gatan Imaging Filter. The sample is cooled with a side-entry liquid helium stage. Further experimental details are discussed in section 8.

## 3. Comparison of Vortex Imaging Techniques

Here, we compare transmission electron microscopy with other techniques used to image flux vortices. As shown by table I, transmission electron microscopy (TEM) offers advantages over other imaging techniques such as magnetic force microscopy, scanning SQUID and Hall probe microscopy, Kerr effect microscopy and the Bitter technique. In principle, the spatial resolution is ~2 nm, limited by the Lorentz lens aberrations, but in practice, it is limited by the signal-to-noise effects discussed in section 7. Unlike all other imaging techniques, electron microscopy is sensitive to the magnetic flux throughout the material and not simply the fields at the surface. A second important advantage is that the vortices can be imaged in real time and thus dynamic studies of individual vortices are possible. It should be noted that magneto-optical imaging can be very fast (time resolution ~$10^{-8}$ s) but individual vortices cannot be seen in this mode of operation. Studies where individual vortices have been observed report a time resolution of 0.1 s [26]. Neutron diffraction is also used extensively in the investigation of flux vortices but has not been included in table I as it is a purely diffractive technique. It is reviewed in ref 27.

The disadvantage of electron microscopy is the high cost (~£1M) of a microscope equipped with a field emission gun, Lorentz lens and a liquid helium cooled specimen stage. However, as demonstrated in this paper, the microscope need not be dedicated to the imaging of vortices and the same microscope we use here has been used for many other purposes such as high resolution imaging, chemical analysis and electron holography of magnetic and semiconducting samples. The liquid helium cooled stage used here is a side entry holder so that a dedicated cryo-electron microscope is not needed.

The field emission gun on a modern commercial electron microscope provides a beam which is sufficiently bright for imaging vortices and the major experimental challenge lies not with the microscope itself but in the preparation of thin (~200 nm), flat samples of superconducting material tens of microns in size with very few defects. Such samples are necessary so that diffraction contrast does not obscure contrast from vortices and their preparation is discussed in detail in section 8.1.

| Technique | Quantitative? | Sensitive to... | Spatial resolution | Maximum field of view | Time resolution | B-field sensitivity |
|---|---|---|---|---|---|---|
| **Fresnel TEM*** | Semi-quantitative | magnetic flux | ~200 nm | 30 μm | Video rate ~1/30 s | ~3×10$^{-4}$ T |
| **TEM with phase plate*** | Semi-quantitative | magnetic flux | ~2 nm | 30 μm | Video rate ~1/30 s | ~0.002 T |
| **Electron holography (TEM)*** | Yes | magnetic flux | ~10 nm | 3 μm | ~0.5 s | ~0.006 T |
| **Bitter patterning** | No | surface field gradient | ~200 nm | no limit | Static | ~10$^{-4}$ T |
| **Magneto-optical imaging** | Yes | surface field gradient | 1 μm | ~mm | 0.1 s (single vortex imaging) | 3×10$^{-6}$ T |
| **MFM** | No | surface field gradient | 100 nm | ~20 μm | ~5 mins | 10$^{-3}$ T |
| **Scanning Hall probe** | Yes | surface field | 1 μm | ~50 μm | seconds | 3×10$^{-6}$ T |
| **Scanning SQUID** | Yes | surface field | 2 μm | ~500 μm | seconds | 10$^{-8}$ T |
| **Scanning Tunnelling Micrscopy** | Yes | surface density of states | ~0.1 nm | ~10 μm | tens of seconds – minutes | N/A |

**Table I:** Comparison of techniques for imaging flux vortices. Most of the techniques measure the B-field associated with the vortices but we have also included scanning tunnelling microscopy which measures the local density of states. The values given here are necessarily approximate as each technique does not measure the same quantity and the most rapid measurements cannot be made at the best spatial resolution but it should enable an overview of the possibilities [26, 28, 29, 30, 31, 32, 33, 34]. *See section 7.

## 4. Orientation of Flux Vortices in Thin Samples

It is not immediately apparent why vortices should penetrate thin film samples at angles close to the plane normal irrespective of the angle at which the B-field is applied. To give some insight, we discuss a simplified model for the competing energies which determine the orientation of a fluxon in a thin sample subject to an applied B-field.

There are two main contributions to the energy: first the self-energy (also called the line-energy or 'tension') which is the amount of energy needed to produce an extra length of vortex. The other contribution expresses the interaction of the vortex with the applied field and is similar to the energy of a magnetic dipole in a field. We discuss each of these contributions in turn.

We can derive the self energy by treating the vortex as a solenoid. The energy of a solenoid of self inductance $L$, carrying a current $I$ is

$$U_{self} = \tfrac{1}{2} LI^2 \tag{14}$$

We now re-express this equation so that it contains terms more familiar to the study of flux vortices.

The self inductance is by definition, the total magnetic flux carried by the solenoid divided by the current.

$$L \equiv \Phi / I \tag{15}$$

Neglecting edge effects for the moment (using the long solenoid approximation), the B-field in a solenoid of length $l$ with $N$ turns can be found using Ampere's law

$$\oint \mathbf{B}.d\mathbf{l} = \mu_0 I_{enclosed} \tag{16}$$

$$B = \mu_0 NI / l \tag{17}$$

If we call the area of the loops of wire in the solenoid $S$, the flux corresponding to the magnetic flux quantum is given by

$$\Phi_0 = \mathbf{B}.\mathbf{S} = \mu_0 NIS / l \tag{18}$$

This allows us to find an expression for the number of loops of wire in the solenoid

$$N = \Phi_0 l / \mu_0 IS \tag{19}$$

The flux referred to by equation (15), however, corresponds to the total flux carried by $N$ loops of wire i.e.

$$\Phi = N\Phi_0 \tag{20}$$

Thus

$$L = \Phi / I = \mu_0 N^2 S / l = \frac{\Phi_0^2 l}{I^2 S} \tag{21}$$

So the energy carried by a flux vortex of length $l$ is then

$$U_{self} = \tfrac{1}{2} LI^2 = \frac{1}{2} \frac{\Phi_0^2 l}{\mu_0 S} \tag{22}$$

Notice the equation is now expressed in terms familiar to the study of vortices and that the number of turns, $N$, which has no physical meaning for a vortex, has cancelled out. The only unknown is the area of the solenoid, $S$. A flux vortex differs from a solenoid because the current is not concentrated at a single radius but decays away to zero with increasing distance from the core. Thus $S$ is an effective area and we would expect $S \sim \pi \Lambda^2$ if we regard the vortex a solenoid with a radius equal to the penetration depth. The textbook formula for a long flux vortex [3] verifies this expectation

$$U_{self} = \frac{1}{2}\frac{\Phi_0^2 l}{\mu_0 (\pi \Lambda^2)}\left(\frac{1}{2}\ln\left(\frac{\Lambda}{\xi}\right)\right) \qquad (23)$$

and allows us to identify the effective area, $S$, of the solenoid as

$$S = \pi \Lambda^2 \left(\frac{2}{\ln(\Lambda/\xi)}\right) \qquad (24)$$

Equation (22) applies only to a long fluxon. In a finite solenoid stray field leaks near the ends, reducing the overall energy since less flux is enclosed by the current. The energy of a finite solenoid can be written as the energy of a long solenoid multiplied by a correction factor, $K(R)$, called the Nagaoka factor [35, 36] as shown in equation (25). A approximation for the Nagaoka factor correct to 6 significant figures which is straightforward to implement is given in ref. 37. The Nagaoka factor depends only on the aspect ratio, $R$ (the ratio of the length of the solenoid to its radius), and is plotted in Figure 4.

$$U_{self} = \tfrac{1}{2}LI^2 = \frac{1}{2}\frac{\Phi_0^2 l}{\mu_0 S}K(R) \qquad (25)$$

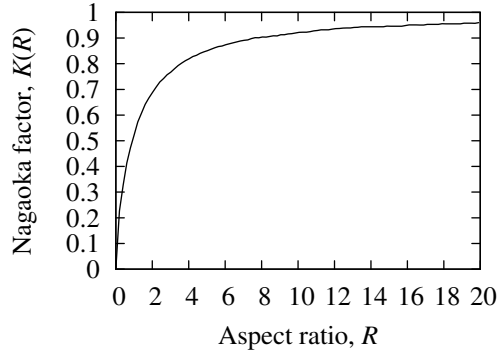

**Figure 4:** Nagaoka Factor, *K(R)*, as a function of aspect ratio, *R* [35, 36].

It should be noted that the boundary conditions for vortices and solenoids are different in that the stray field from a flux vortex cannot re-enter the superconductor. In contrast to a solenoid where the circulating current is concentrated at a single radius, the current surrounding the vortex core is distributed in such a way that the B-field generated does not re-enter the superconductor. The energy of the solenoid depends on the amount of magnetic flux enclosed within the circulating current not on what the field does outside the vortex and so we do not expect this distinction to greatly alter our argument which is to demonstrate that the vortex is likely to be oriented within a few degrees of the specimen normal. If anything, we shall underestimate the flux enclosed and the vortex will have a higher self energy than the solenoid so that our estimates of the angle will be overestimates.

We now examine the second contribution to the energy: the interaction of the external the B-field with the vortex. The external B-field affects the monopoles at the top and bottom of the vortex and the interaction is that of a magnetic dipole **μ** in an external field **B** and the associated energy is

$$U_{field} = -\boldsymbol{\mu} \cdot \mathbf{B} \tag{26}$$

To evaluate the dipole moment of a flux vortex, we again treat it as a solenoid with an effective area *S*.

The magnetic dipole moment of a single loop of wire carrying a current *I* is *IS* so the dipole moment of the solenoid is *N* of these loops added together.

$$\mu = ISN \tag{27}$$

Using equation (18), the magnetic moment can be rewritten in terms of the flux quantum

$$\mu = \Phi_0 l / \mu_0 \tag{28}$$

So the energy of the fluxon in the external B-field is given by

$$U_{field} = -\boldsymbol{\Phi_0} \cdot \mathbf{B} l / \mu_0 \tag{29}$$

Finally, we have the total energy of a flux vortex in a B-field:

$$U = U_{self} + U_{field} = \frac{1}{2} \frac{\Phi_0^2 l}{\mu_0 S} K(R) - \frac{\boldsymbol{\Phi_0} \cdot \mathbf{B} l}{\mu_0} \tag{30}$$

Figure 5 shows the geometry of a typical sample for electron microscopy which is a thin film of thickness *h* and is taken to be infinitely wide in the other directions. When a flux vortex is tilted at an angle $\beta$ with respect to the specimen surface normal, the length of the vortex is $h/\cos\beta$ and if a B-field is applied at an angle $\beta_B$ to the surface normal, the energy is

$$U = \frac{1}{2} \frac{\Phi_0^2}{\mu_0 S} \frac{h}{\cos\beta} K\left(\frac{R_0}{\cos\beta}\right) - \frac{\Phi_0 B \cos(\beta_B - \beta)}{\mu_0} \frac{h}{\cos\beta} \tag{31}$$

where $R_0$ is the aspect ratio of the vortex when it is vertical.

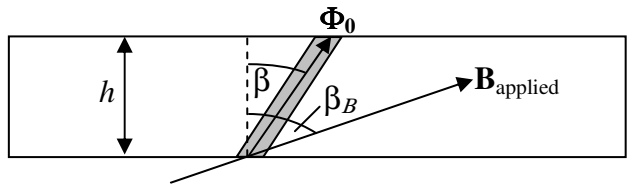

**Figure 5:** Geometry for a Fluxon in an Applied B-field in a thin foil sample.

To find the equilibrium orientation, we numerically minimised the total energy with respect to $\beta$, the orientation of the vortex. The results are plotted in Figure 6 for an applied field of 30 G (the largest field applied in the experiments performed here). In the simulations here and throughout the paper, we compare Nb and BSCCO ($Bi_2Sr_2CaCu_2O_{8-\delta}$) as Nb has fairly narrow vortices (we used $\xi = 39$ nm, $\Lambda = 52$ nm [3]) and BSCCO has very wide vortices ($\xi = 3$ nm, $\Lambda = 300$ nm [8]) so we consider two extreme cases.

The plots demonstrate that for specimen thickness in the range 100-500 nm (typical in these experiments), vortices in Nb tilt by less than 14° away from the sample normal even when the field is applied along the sample surface. According to these calculations, BSCCO gives larger tilt angles up to 33° as its line energy is smaller. However, this analysis assumed that BSCCO had conventional vortices which is far from being the case and its behaviour is quite different to that predicted here as discussed at the end of this section. The simulation results are included here for a comparison of wide, conventional vortices with the narrower vortices of Nb.

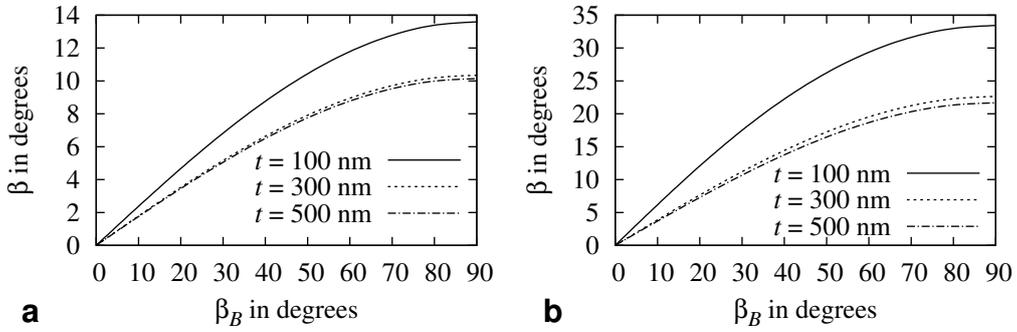

**Figure 6:** Orientations of Flux Vortices. (a) the tilt angle $\beta$ adopted by the flux vortex for various angles ($\beta_B$) of an applied field of 30 G in Nb for different specimen thicknesses. (b) the same for BSCCO. Note that the behaviour of real BSCCO is quite different from that predicted here as explained in the text.

An aspect neglected in this analysis is the boundary condition that the current loops of the solenoid must be parallel to the superconductor at the surface implying that the field near the centre of the vortex must be locally normal to the surface. In the limit of an infinitesimally thin vortex this would constrain the field to go normally through the film. For thicker specimens, the vortex may be tilted within the sample but then must bend round to be normal near the surface. This will tend to reduce the tilting angle so the simple analysis presented here should be regarded as an *upper* limit on the angle adopted by conventional vortices. There is a surprisingly small amount of literature on the subject of fluxon orientation in thin superconductors. Two papers which deal with the subject are refs. 38 and 39 where London theory is used to give the B-field distribution due to the vortices, the free energy is calculated from this and then minimised to give the equilibrium vortex shape. In all simulations of the appearance of vortices in electron micrographs to date, the tilt and shape of the vortices is put in by hand. It would be advantageous to combine the energy minimisation approach with the image simulations to produce images of vortices in their equilibrium shape and we are currently working on this.

It should also be noted that for tilted vortices in thin samples, the direction of the vortex core can differ substantially from the direction of the B-field lines [40, 41]. In this simplified model the vortex core is entirely neglected as it makes a small change to the energy and the "orientation of the solenoid" should be interpreted as the direction the B-field lines take near the centre of the vortex. It is the orientation of the field lines which determine the appearance of an electron microscopy image rather than the orientation of the core.

An aspect not considered in the model is a possibility that can occur in layered superconductors which consist of atomic planes of superconducting material separated by layers where the number density of Cooper pairs is greatly reduced. Vortices which occur on the superconducting planes have supercurrents confined to these planes and are termed "pancake" vortices. When a field is applied normal to the superconducting layers, it penetrates via a stack of pancake vortices. When the applied field is tilted by a small angle with respect to the superconducting layers, the stack of pancakes first tilts and the situation is similar to the tilting of conventional vortices. As the tilt angle increases, it can be energetically favourable for the stack of pancake vortices to break as illustrated in Figure 7(a). The penetrating flux then generates a sequence of alternating pancake vortices and Josephson vortices. Josephson vortices consist of currents in the superconducting planes connected by tunnelling currents (indicated by dashed lines) which traverse the normal regions between the layers. The tunnelling current is generated by the same mechanism as the tunnelling current in a Josephson junction [42].

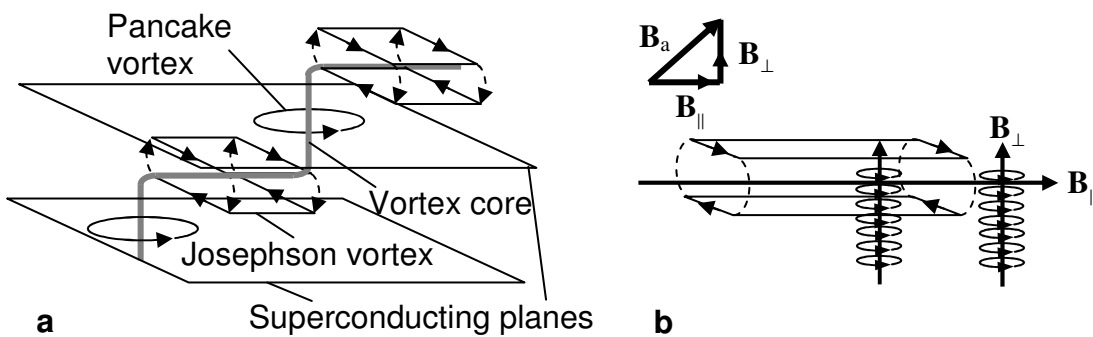

**Figure 7:** Vortices in layered superconductors. Solid lines with arrows show supercurrents and dashed lines show tunnelling currents. (a) Vortex Kinking: the vortex is a staircase composed of pancake vortices and Josephson vortices. The current loops surrounding the vortex core are not to scale. In reality they extend for hundreds of nanometres whereas the spacing of the superconducting planes is ~1 nm. (b) Crossing Lattice: the component of the field parallel to the superconducting planes ($B_\parallel$) penetrates via Josphson vortices and the component normal to the superconducting planes ($B_\perp$) penetrates via stacks of pancake vortices.

The cores of Josephson and pancake vortices are very small: on the order of the interlayer spacing (~nm) but the surrounding supercurrents and the associated B-fields are on a much larger scale of hundreds of nanometers (note that Figure 7(a) is not to scale). The B-field does not follow the highly kinked path of the core but follows the average tilt angle.

Josephson vortices have a much lower line energy than conventional vortices because the width of the surrounding current is so great (recall that the line energy is inversely proportional to the solenoid area–equation (25)). The height and width of the current surrounding a Josephson vortex are determined by the penetration depths parallel and perpendicular to the superconducting planes which in YBCO ($YBa_2Cu_3O_{7-\delta}$) are 200 nm and 1 µm respectively [43]. The lower line energy of Josephson vortices means that much larger average tilts are possible and Beleggia *et al.* [43] have simulated Fresnel images of kinked vortices for comparison with experimental images of vortices in YBCO and find that the two compare best when average tilt angles from 70–85° are used at fields of only 3 G.

There is another possibility illustrated in Figure 7(b) which is the formation of a crossing lattice where the horizontal component of the field induces Josephson vortices and the vertical component induces pancake vortices but the two coexist separately. This scenario occurs in very highly anisotropic superconductors like BSCCO. Unlike the case of vortex kinking, a crossing lattice always forms and there is no tilting of the pancake stacks at any field [44]. The images shown later in this paper are from superconducting BSCCO and the vortices observed are stacks of pancake vortices lying normal to the superconducting layers. Note that BSCCO cleaves along its superconducting layers so for the experiments carried out here, the thin film surface is parallel to the superconducting layers and the stacks of pancake vortices are expected to be normal to the specimen surface.

Josephson vortices are even harder to observe than conventional vortices using transmission electron microscopy. They are so wide that the deflection of the electron beam is very small ($10^{-7}$ rad) and so far, Josephson vortices have not been imaged directly although a recent theoretical paper indicates that it should be just possible [45].

## 5. Assessment of the Advantages of using a Higher Acceleration Voltage

One of the main difficulties with using transmission electron microscopy to image flux vortices is the requirement to work with thick specimens to maximise the phase shift of the electrons. The reason that thick superconducting specimens give a larger phase shift is rather different to the situation with ferromagnetic samples. The thicker a ferromagnetic sample, the more magnetic material is traversed by the electrons and the larger the phase shift. The width of the flux vortex, however is determined by the penetration depth and not the sample thickness. The reason that it is advantageous to image flux vortices in thick samples is that the phase shift from the B-field within the specimen is partially cancelled by the B-field from the stray field which emanates from points at which the fluxon touches the top and bottom of the sample. The thicker the sample, the further apart these points are and the weaker the component of the stray field normal to the electron beam over the length of the vortex as illustrated in Figure 8.

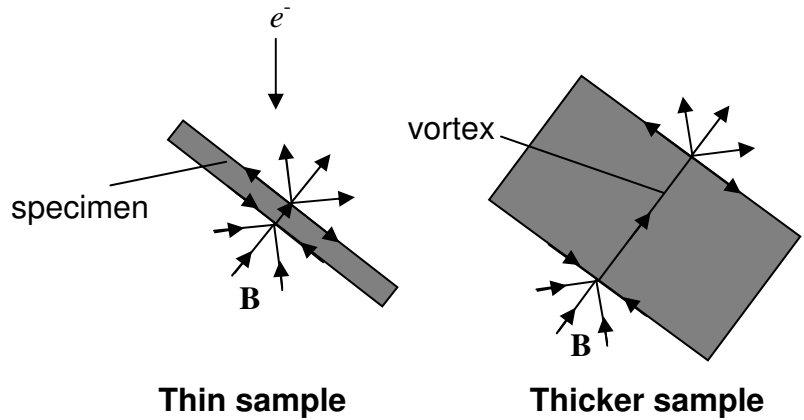

**Figure 8:** Schematic diagram illustrating the B-field from thin and thick samples.

Specimens are typically 200–500 nm thick which is considerably thicker than conventional microscopy specimens which are < 10 nm. This has the obvious disadvantage that the images appear very dark and long exposures (~20 s) are needed to image the vortices. To improve the situation, Tonomura *et al.* [46] developed the 1 MV electron microscope with a field emission gun. We now assess the improvement that comes from using a higher voltage.

Increasing the acceleration voltage has four main benefits. First, the brightness of the electron gun is increased so that the current density in the image is higher; secondly the fraction of electrons which are 'absorbed' by a given specimen thickness is reduced; thirdly, the fraction of inelastically scattered electrons which contribute to the image is reduced and finally, the electrostatic contribution to the phase shift of the electron wave is reduced but the magnetic contribution remains the same. We discuss the contribution of each in turn. It should be noted that as fluxons are very large objects compared with the atomic scale structures which are normally investigated by electron microscopy, the effect of lens aberrations is quite insignificant and thus they are not discussed here.

*5.1 Gun Brightness*

The maximum achievable gun brightness is limited by the inevitable transverse momentum of electrons coming from the gun. A higher acceleration voltage reduces the ratio of the transverse to axial momenta of the electrons so that a higher brightness can be achieved by increasing the acceleration voltage. The 1 MV electron microscope used by the Tonomura group has a gun brightness of $2 \times 10^{10}$ Acm$^{-2}$sr$^{-1}$: an order of magnitude greater than that of our microscope. If the acceptance angles are similar, we expect the current density in the image to be increased by a similar amount.

*5.2 Absorption*

Absorption is a term used in microscopy to mean any electrons which interact with the specimen in such a way that they do not contribute to the image. These are mainly electrons which are scattered to such high angles by the specimen that they are intercepted by aperatures: electrons which are genuinely absorbed by the specimen are comparatively rare.

If there is an objective aperture of semi-angle $\varepsilon$, we can estimate the number of electrons scattered outside the aperture by considering the elastic scattering cross section. Egerton [47] gives the following formula for the ratio of the scattering cross-section ($s$) for angles from 0 to $\varepsilon$ divided by the total cross-section.

$$\frac{s_e(0 \to \varepsilon)}{s_e} \approx \left(1 + \frac{a_0^2}{\varepsilon^2 \lambda^2 Z^{2/3}}\right)^{-1} \tag{32}$$

where $a_0$ is the Bohr radius and $Z$ the atomic number.

The total cross section can be estimated using Langmore's formula [47].

$$s_e = \frac{(1.5 \times 10^{-24} \text{ m}^2) Z^{3/2}}{(v/c)^2}\left[1 - \frac{Z}{596(v/c)}\right] \tag{33}$$

These cross sections can be converted to mean free paths $d$ using

$$d = \frac{1}{n_a s} \tag{34}$$

where $n_a$ is the number density of atoms.

After passing through the specimen, a fraction of the electrons is lost due to scattering outside the objective aperture. The fraction of the incident electrons which is not scattered to such high angles and thus contributes to the image on traversing a specimen thickness of thickness $t$ is then

$$f(0 \to \varepsilon) = \exp(-t/d_\varepsilon) \tag{35}$$

where $d_\varepsilon$ is the mean free path appropriate for elastic scattering between angles 0 and $\varepsilon$.

For Nb ($Z = 41$), these formulae give a total elastic mean free path of 30 nm at 300 kV and 44 nm at 1 MV. For an objective aperture of size 10 mrad, $d_\varepsilon = 48$ nm at 300kV and 181 nm at 1MV. For a 200 nm thick Nb specimen, the fraction of electrons which pass through a 10 mrad objective aperture and contribute to the image is 1.6% for 300 kV electrons and 33% for 1 MV. For a 500 nm thick specimen only 0.003% pass through at 300 kV and 6% at 1 MV.

*5.3    Inelastic Scattering*

In a similar manner we can calculate the ratio of elastically scattered to inelastically scattered electrons in an image. Scattering from vortices is an elastic process and inelastically scattered electrons will tend to obscure the vortices. After traversing a specimen of thickness $t$, the fraction of electrons which have not been inelastically scattered is given by [47]

$$f_0 = \exp(-t/d_{inelastic}) \tag{36}$$

where $t$ is the specimen thickness and $d_{inelastic}$ the inelastic mean free path. The inelastic mean free path is given approximately by

$$d_{inelastic} = \frac{106(E_0/E_m)}{\ln(2\varepsilon E_0/E_m)} \frac{1+E_0/1022}{(1+E_0/511)^2} \tag{37}$$

where $\varepsilon$ is the collection semi angle of the objective lens in mrad, $E_0$ the incident beam energy in eV and $E_m$ is a constant dependent on atomic number given in eV by $E_m \approx 7.6Z^{0.36}$ [47].

According to this formula, for Nb at an acceleration voltage of 300 kV, the inelastic mean free path for a collection semi-angle of 10 mrad is 106 nm and at 1 MV it is 127 nm. Thus the ratio of elastically scattered electrons to the total number is 15% at 300 kV compared with 21% at 1 MV for a 200 nm thick specimen and 0.8% at 300 kV compared with 1.9% at 1 MV for a 500 nm thick specimen.

It should be noted that our microscope is equipped with an energy filter and consequently it is possible to form images where the fraction of elastically scattered electrons is considerably higher than the figures quoted here.

*5.4    Phase Shift*

The phase shift of an electron travelling along the $z$ axis after passing through a specimen (see Figure 9 for geometry) with a mean inner potential $V_0$ and producing a B-field **B** is given by:

$$\phi = C_E \int V_0 dz - \frac{2\pi e}{h} \int \mathbf{B}.d\mathbf{S} \tag{38}$$

where $C_E$ is a constant determined solely by the acceleration voltage of the microscope $V$ and given by:

$$C_E = \frac{2\pi e}{h} \sqrt{\left(2meV\left(1+\frac{eV}{2mc^2}\right)\right)} \left(\frac{eV+mc^2}{eV(eV+2mc^2)}\right) \tag{39}$$

where $m$ is the rest mass of the electron, $e$ the electron charge and $c$ the speed of light in a vacuum.

It can be seen that increasing the acceleration voltage does not affect the magnetic contribution to the phase shift but reduces the electrostatic component. With flux vortices, the magnetic contribution is the only component of interest so it is advantageous to reduce the electrostatic contribution by increasing the voltage. At 300 kV, $C_E = 6.52 \times 10^6$ m$^{-1}$V$^{-1}$ and at 1MV, $C_E = 5.38 \times 10^6$ m$^{-1}$V$^{-1}$ so $C_E$ is 82% of its former value.

*5.5  Conclusion*

We conclude that the greatest advantage of increasing the voltage is that fewer electrons are absorbed. For a 200 nm thick Nb sample, about 20 times fewer electrons are absorbed using a 1 MV microscope compared with 300 kV and this factor increases exponentially with increasing specimen thickness. In addition, the increased source brightness increases the current density at the image by a further factor of 10. In comparison, the reduction in inelastic scattering by increasing the voltage is nullified by the use of an energy filter and the reduction of the electrostatic component of the phase shift is modest.

Increasing the number of electrons which contribute to the image will make the image appear brighter (the current density would be increased by a factor of ~200 for 200 nm thick Nb according the calculations here) and will also reduce diffraction contrast so that the specimen would appear thinner. It should be noted that the 'absorption' could be reduced without increasing the acceleration voltage by making the objective aperture wider but this would increase the effect of lens aberrations in the image. In microscopes with aberration corrected objective lenses, the specimen appears thinner as larger apertures can be used. We are not aware of any plans to make an electron microscope with an aberration corrected Lorentz lens but one advantage of such a microscope would be its ability to take images from thicker samples.

## 6.  Calculation of Phase Shift due to a Flux Vortex

Here we assess the optimum conditions for imaging flux vortices in superconductors by simulating the appearance of images observed in the transmission electron microscope. The flux vortices affect only the phase of the electron wavefunction and so we begin with an explanation of how the phase can be calculated. Following this, we simulate images and assess how they may be optimised. The calculations of the phase shift from flux vortices are based on the equations derived by Pozzi and coworkers at the University of Bologna and are rederived here for completeness. The original papers from which the equations come are referenced in the text. As discussed in section 4, flux vortices are expected to lie within a few degrees of the specimen plane normal and the vortices simulated here are assumed to lie normal to the sample surface.

*6.1  Phase Shift from a Vortex of Infinitesimal Radius*

When an electron wave passes through a sample, its amplitude is altered due to absorption and its phase is changed by interactions with electric and magnetic fields associated with the sample. Taking the electron beam to be travelling in the *z* direction, the phase shift of an electron is given by:

$$\phi(x,y) = C_E \int_\Gamma V(x,y,z)dz - \frac{2\pi e}{h}\int_S \mathbf{B}(x,y,z).\mathbf{dS} \qquad (40)$$

where $V$ is the electric potential and **B** is the flux density. $\Gamma$ is an infinitely long, straight line trajectory parallel to $z$ passing through a point $(x, y)$ in the specimen plane. The phase has an arbitrary zero point and we define a particular reference trajectory $\Gamma_0$ to represent zero phase. The vector area S in the second term is the area formed by joining $\Gamma$ and $\Gamma_0$ infinitely far from the specimen as illustrated in Figure 9. Note that this implies that only the components of **B** normal to **S** contribute to the phase. As flux vortices are purely magnetic, we shall only calculate the contribution to the phase from the second term.

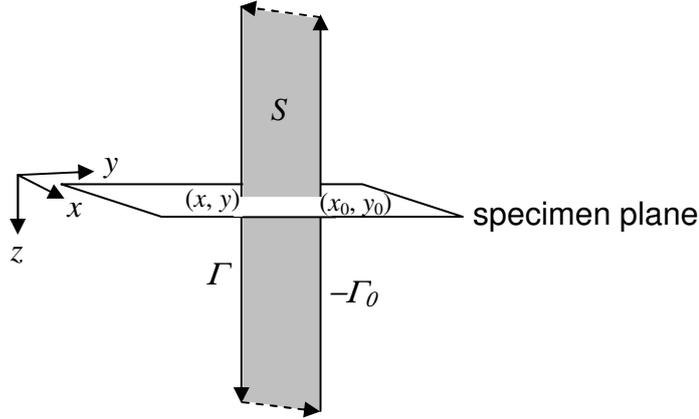

**Figure 9:** Diagram showing how the trajectory $\Gamma$ and the reference trajectory defining zero phase $\Gamma_0$ can be connected (dashed arrows) infinitely far from the sample to form a closed circuit with a bounded area of $S$.

The flux density **B** from a flux vortex can be found by first considering a vortex of infinitesimal radius. For simplicity, we consider the case of a fluxon lying normal to the electron beam direction, pointing in the *x* direction [48] as shown in Figure 10. (In Appendix 1 we consider the case of a vortex tilted by an angle $\alpha$ about *y*.) Placing the origin, O, at the point at which the vortex reaches the sample surface, we initially consider an infinitely long vortex in a semi-infinite specimen which takes up the whole half-plane $x < 0$. In this case, the B-field in the vacuum must emanate from the point at which the fluxon reaches the surface, die away to zero infinitely far from the fluxon and not re-enter the diamagnetic sample. It can be seen that to obey these boundary conditions, the flux in the vacuum is equivalent to field from a monopole placed at the origin but zero inside the diamagnetic specimen apart from at the fluxon itself.

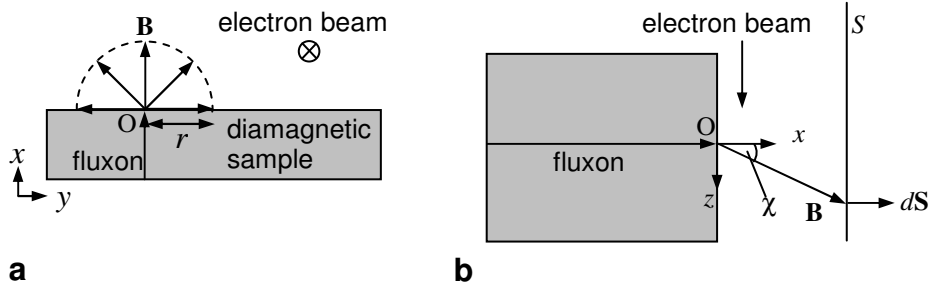

**Figure 10:** (a) The fluxon viewed along the direction of the electron beam. The magnetic flux density **B** in the vacuum from a fluxon of infinitesimal radius can be found by integrating over the dashed hemisphere. (b) Geometry in the *xz* plane, viewing the specimen in cross-section.

The field in the vacuum from the fluxon can be found using the fact that the magnetic flux inside the fluxon must be equal to the flux that crosses the dashed hemisphere shown in Figure 10(a).

$$\int_{\text{hemisphere}} \mathbf{B}_{\text{vacuum}} \cdot d\mathbf{S} = \Phi_0 \Rightarrow \mathbf{B}_{\text{vacuum}} = \frac{\Phi_0}{2\pi r^2}\hat{\mathbf{r}} \qquad (41)$$

where **r** is the radius vector measured from the origin and $\hat{\mathbf{r}}$ is a unit vector pointing in the same direction.

Knowing the flux density in the vacuum, the magnetic contribution to the phase shift in the vacuum can be calculated using equation (40) and referring to the geometry in Figure 10(b),

$$\phi_{\text{vacuum}} = -\frac{2\pi e}{h}\int \mathbf{B}.d\mathbf{S} = -\frac{2\pi e}{h}\int \frac{\Phi_0 dS \cos\chi}{2\pi r^2} \qquad (42)$$

However, since $\Phi_0 = h/2e$ and $dS\cos\chi/2\pi r^2$ is simply the element of solid angle $d\Omega$ subtended by the area element $dS$, the phase shift is given by

$$\phi_{\text{vacuum}} = -\frac{1}{2}\Omega \qquad (43)$$

This result also holds for the inclined fluxon (See Appendix 1) but as an illustrative example, we continue the analysis of a fluxon lying normal to the electron beam. From Figure 11, it can be seen that the phase in the vacuum is dependent only on the azimuthal angle γ and is given by:

$$\phi_{\text{vacuum}} = -\frac{1}{2}\Omega = -\frac{1}{2}\cdot\frac{\gamma}{2\pi}4\pi = -\gamma \qquad (44)$$

The phase in the sample is constant as there is no magnetic field inside apart from the fluxon which causes a sudden jump in phase of

$$\Delta\phi_{fluxon} = -\frac{2\pi e}{h}\int \mathbf{B}.d\mathbf{S} = -\frac{2\pi e}{h}\Phi_0 = -\pi \quad (45)$$

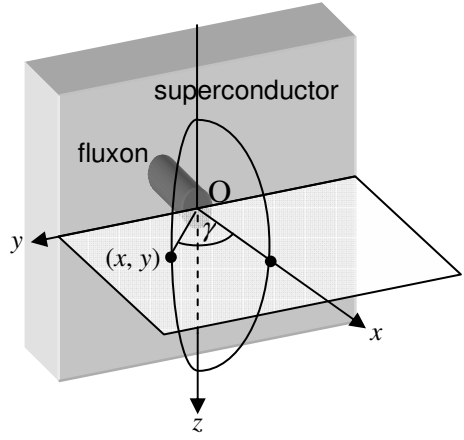

**Figure 11:** Spherical coordinates used to calculate the phase. As the phase depends only on the azimuthal angle γ, the *xz* plane is chosen to define zero phase and the two arcs show the limits of the solid angle used to calculate the phase at a point (*x*, *y*).

This phase shift is shown in Figure 12(b). Phase images are frequently displayed as the cosine of some multiple of the phase partly for historical reasons: when electron holograms were reconstructed optically, the resulting image was the cosine of a multiple of the phase. (Nowadays, a digital reconstruction of a hologram gives the phase itself.) This produces a contour map of the phase and such contour maps resemble magnetic field lines. The cosine of 8 times the phase shift is shown in Figure 12(c). It should be noted that the jump in phase in the specimen exactly compensates for the phase change in the vacuum so that the phase is single valued.

It is interesting to note that rather than splitting the solution into a part which applies in the specimen and a part which applies in the vacuum, the complete solution can be written as

$$\phi = -\frac{1}{2}\left(\gamma + \sin^{-1}(\sin\gamma)\right) = \begin{array}{l} \pi/2 \text{ for } -\pi < \gamma < -\pi/2 \\ -\gamma \text{ for } -\pi/2 < \gamma < \pi/2 \\ -\pi/2 \text{ for } \pi/2 < \gamma < \pi \end{array} \quad (46)$$

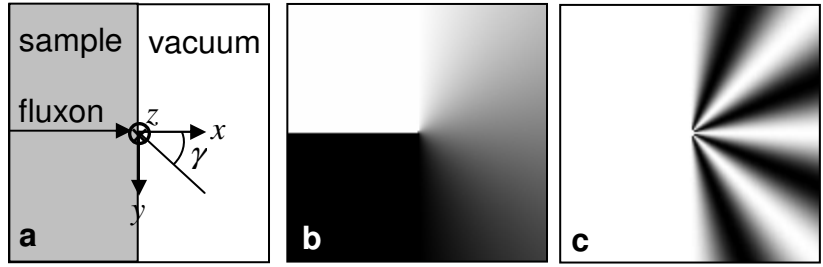

**Figure 12:** (a) Geometry, (b) the phase $\phi$ and (c) $\cos(8\phi)$ for a fluxon of infinitesimal radius.

For a specimen of finite thickness where field enters at one side and exits at the other, the phase shift can be written as

$$\phi = -\frac{1}{2}\left(\gamma_u + \sin^{-1}(\sin \gamma_u)\right) + \frac{1}{2}\left(\gamma_l - \sin^{-1}(\sin \gamma_l)\right) \qquad (47)$$

where $\gamma_u$ and $\gamma_l$ are measured from origins $O_u$ and $O_l$ as shown in Figure 13. The first term in equation (47) is the same as that for a vortex in a semi-infinite specimen, producing a phase jump in the specimen and an angular phase ramp in the vacuum at the right hand side of the specimen. The second term produces an opposite angular phase ramp in the vacuum at the left hand side of the specimen and also contains a phase jump in the vacuum which cancels the phase jump from the first term. The resulting phase shift is shown in Figure 13(b). Figure 13(c) shows the cosine of 8 times the phase with arrows superimposed to show the direction of the projected B-field.

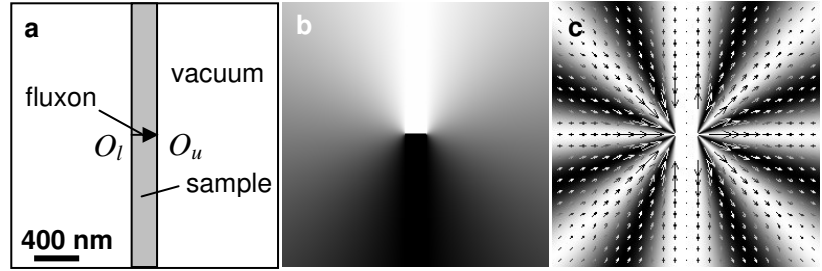

**Figure 13:** 200nm long fluxon of infinitesimal radius described by equation (47). (b) shows the phase shift which ranges from π/2 (white) to –π/2 (black). (c) shows the cosine of 8 times the phase with arrows superimposed to indicate the direction of the projected B-field.

Remarkably, the expression for the phase shift (derived in appendix 1) when the specimen normal is tilted at an angle $\alpha$ (see Figure 2) to the electron beam is similarly straightforward [49].

$$\phi = -\frac{1}{2}\left(\gamma_u + \sin^{-1}(\sin \alpha \sin \gamma_u)\right) + \frac{1}{2}\left(\gamma_l - \sin^{-1}(\sin \alpha \sin \gamma_l)\right) \qquad (48)$$

The effect on the phase of tilting the specimen is shown in Figure 14.

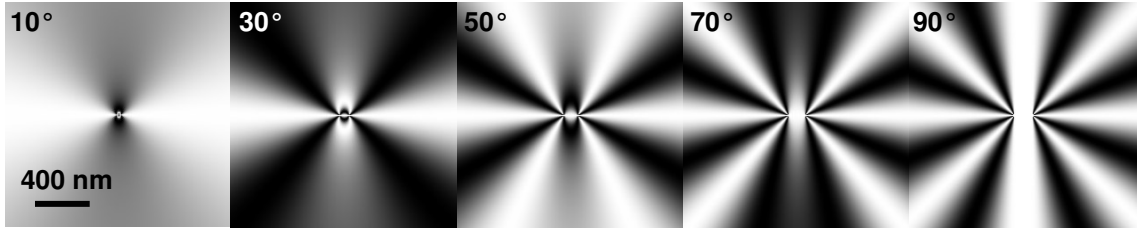

**Figure 14:** 200nm long, fluxon of infinitesimal radius. The images show the cosine of 8 times the phase shift at various tilt angles ($\alpha$).

It should be noted that this treatment can be extended [50] to deal with the case of vortices which take an arbitrary path through the superconductor rather than the straight line path described here.

*6.2 Flux Vortex with a Finite Radius*

So far, we have treated the case of a fluxon with an infinitesimal radius. A fluxon with a finite radius can be modelled as a sum of infinitesimal fluxons since phases are additive [49]. In mathematical terms, the infinitesimal fluxon is convolved with a shape function $s(u)$ that represents the internal structure of the vortex ($u$ is the radius measured from the central axis of the vortex).

For conventional type II superconductors, the Clem model [51] gives the B-field as a function of the radius from the centre of the vortex $u$ as:

$$B = \frac{\Phi_0}{2\pi\Lambda\xi_V} \frac{K_0\left(\sqrt{u^2 + \xi_V^2}/\Lambda\right)}{K_1(\xi_V/\Lambda)} \tag{49}$$

where $\Lambda$ is the penetration depth and $\xi_V$ is a variational parameter which has the same order of magnitude as the coherence length, $\xi$, and in what follows we use the coherence length as the value of $\xi_V$. $K_0$ and $K_1$ are first and second order hyperbolic Bessel functions. The B-field profiles for Nb and BSCCO are shown in Figure 15.

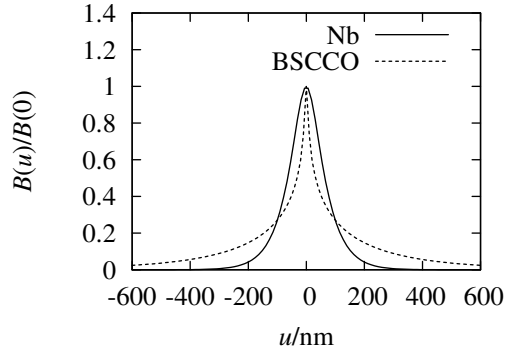

**Figure 15:** The flux density profile for Clem-type vortices in niobium using $\Lambda$ = 52 nm and $\xi_V$ = 39 nm and in BSCCO using $\Lambda$ = 300 nm and $\xi_V$ = 3 nm.

Thus the shape function is $s(u) = \mathcal{N} K_0 \left( \sqrt{u^2 + \xi_v^2} / \Lambda \right)$ where $\mathcal{N}$ is a normalisation constant chosen so that when the shape function is integrated over the plane normal to the fluxon, it gives unity. The zeroth-order hyperbolic Bessel function was evaluated using the method described in *Numerical Recipies* [52]. It is more convenient to find $\mathcal{N}$ by numerically normalising $s(u)$ than to evaluate it analytically as this avoids the need to evaluate $K_1$. Unless the fluxon is parallel to the electron beam, the radius $u$ will not lie in the *xy* plane. Thus it is necessary to project the shape function to account for the specimen tilt before performing the convolution. The procedure is shown in Figure 16.

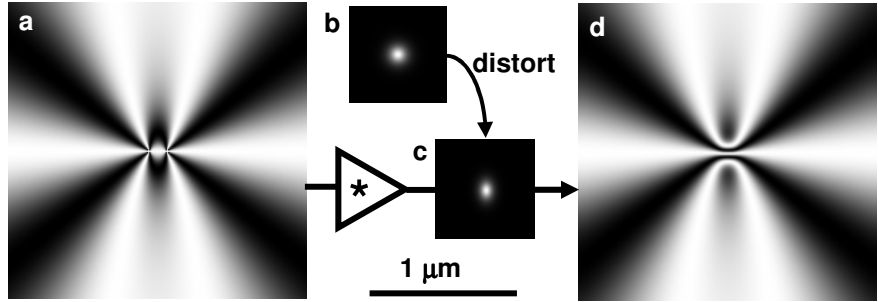

**Figure 16:** (a) $\cos(8\phi)$ for a 200 nm long fluxon of infinitesimal radius tilted at 45° with respect to the electron beam. (b) Clem-type shape function $s(u)$ with $\Lambda = 52$ nm and $\xi = 39$ nm viewed along the axis of the fluxon. This is then distorted to account for the 45° tilt to give (c). Convolving (a) with (c) gives the result for a finite fluxon shown in (d), also shown as $\cos(8\phi)$.

*6.3    Fourier Space Approach*

The simulation procedures outlined here suffer from the drawback that they do not account for the spreading of the B-field of a vortex as it approaches the surface of the specimen. Such a spreading is inevitable as can be seen from Figure 17 and a model where this does not occur is unphysical. In 1998, Bonevich *et al.* introduced a real-space model which correctly accounts for this boundary condition [53] and in 2001, Beleggia and Pozzi introduced a more elegant Fourier-space approach [54] which has been used in all subsequent simulations performed by that group. In this paper, we use the simpler but less precise real-space method described in section 6.2 as this is quite adequate for a discussion of how to optimise the images from flux vortices although in future we intend to adopt the Fourier space approach. It is also easier to model pancake vortices [54] and Josephson vortices [45] within the Fourier space formalism.

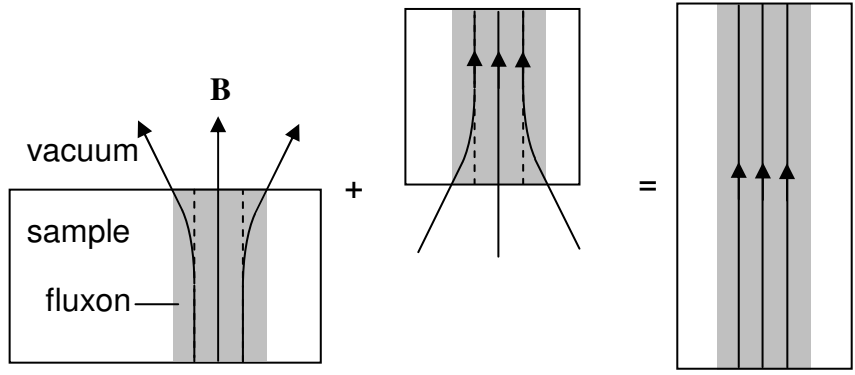

**Figure 17:** B-field spreading near the sample surface. (a) shows a long flux vortex reaching the sample surface. (b) illustrates the same situation but with the B-field entering the sample surface. (a) and (b) must sum to give (c), a long vortex with no surface. This demonstrates that at the sample surface, the flux density must be half that of the bulk and so the spreading of the B-field near surfaces is inevitable.

## 7. Comparison of Different Imaging Techniques

Here we use computer simulation to compare the different techniques for imaging flux vortices using transmission electron microscopy in order to assess which is best for a given situation. We simulate diffraction patterns from vortices as well as holograms, Lorentz images, Foucault images and images obtained with the use of a phase plate and estimate the likely spatial resolution and B-field sensitivity of each.

### *7.1 Deflection of Electrons due to Flux Vortices*

We begin by calculating an approximate upper limit on the deflection angles for electron scattering from a single flux vortex to give a physical insight into the problem and an idea of the range of the diffraction angles involved. In the next section we simulate realistic diffraction patterns. Here, we assume that the B-field is uniform within the fluxon over a radius $W$, neglect the stray field and consider only electrons which travel through the centre of the vortex as these experience the highest deflection. The sample is taken as a thin sheet of thickness $l$ tilted at an angle $\alpha$ to the horizontal and we assume that the vortex is normal to the sample surface as shown in Figure 18(a). If the electrons impinge on the sample with velocity $\mathbf{v}$, those which pass through the B-field experience a Lorentz force $\mathbf{F} = -e\mathbf{v} \times \mathbf{B}$ and are scattered through a small angle $\Theta$.

The magnitude of the Lorentz force is $F = evB_\perp$ where $B_\perp$ is the component of the B-field normal to the electron beam so if the electrons pass through the B-field in a time $T$, they acquire a sideways momentum

$$p_\perp = evB_\perp T \qquad (50)$$

If the B-field through which the electron beam passes persists over a thickness $t$ parallel to the beam, $T = t/v$ and so

$$p_\perp = evB_\perp \frac{t}{v} = eB_\perp t \qquad (51)$$

The small deflection angle is then the ratio of the sideways and vertical momenta

$$\Theta = \frac{p_\perp}{p_\parallel} = \frac{e\lambda B_\perp t}{h}. \qquad (52)$$

From the assumption that the B-field in the vortex is uniform, we take

$$B_\perp = \frac{\Phi_0}{\pi W^2} \sin\alpha = \frac{h}{2\pi e W^2} \sin\alpha. \qquad (53)$$

The thickness, $t$, over which the electron beam experiences a B-field can either be limited by the specimen thickness as occurs at low tilt angles $\alpha$ or limited by the width of the vortex for high tilt angles as illustrated in Figure 18(b). The critical angle $\alpha_{crit}$ at which the cross-over occurs is when

$$\frac{l}{\cos\alpha} = \frac{2W}{\sin\alpha} \quad \text{i.e.} \quad \alpha_{crit} = \tan^{-1}\left(\frac{2W}{l}\right) \qquad (54)$$

The deflection angle is then given by

$$\Theta = \frac{\lambda l}{2\pi W^2} \tan\alpha \quad \text{if} \ \alpha \leq \tan^{-1}\left(\frac{2W}{l}\right) \qquad (55)$$

$$\Theta = \frac{\lambda}{\pi W} \quad \text{if} \ \alpha > \tan^{-1}\left(\frac{2W}{l}\right) \qquad (56)$$

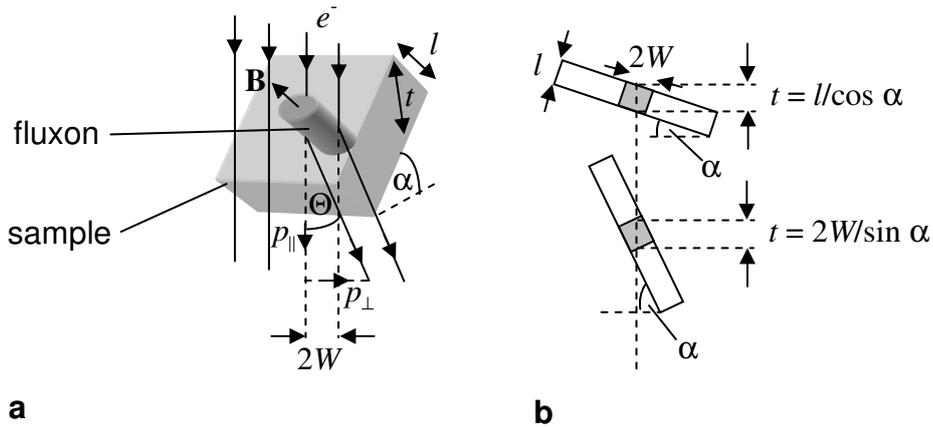

**Figure 18:** (a) Simple model to explain the appearance of Fresnel images. See text for details. (b) The thickness over which the electron beam experiences a B-field is limited at low angles $\alpha < \alpha_{crit}$ by the specimen thickness (upper panel) and at high angles $\alpha > \alpha_{crit}$ by the width of the vortex (lower panel).

Based on equation (24) we take $W = \Lambda\sqrt{2/\ln(\Lambda/\xi)}$ so that for a specimen tilted to $\alpha = 50°$, the projected B-field is 270 G in Nb (using $\Lambda = 52$ nm, $\xi = 39$ nm) and 130 G in BSCCO (using $\Lambda = 300$ nm, $\xi = 3$ nm). The critical angle above which $t$ is determined by the vortex width rather than the specimen thickness is $\alpha_{crit} = 54°$ for Nb and 63° for BSCCO in a 200 nm thick specimen. For the same specimen illuminated by 300 kV electrons and tilted at $\alpha = 50°$, the approximate upper limit on the diffraction angle is then $\Theta = 4$ μrad for Nb and $\Theta = 2$ μrad for BSCCO.

The reader may be surprised that the diffraction angle in equation (56) relevant for high specimen tilt angles $\alpha > \alpha_{crit}$ does not involve the tilt angle $\alpha$. This is because $t$ is proportional to $1/\sin\alpha$ but the projected B-field is proportional to $\sin\alpha$. In a realistic situation simulated in the next section, there is no critical angle beyond which the diffraction pattern remains unchanged as this is an effect caused by neglecting the stray field. The stray field tends to counteract the deflection from the vortex and so it is always advantageous to tilt the sample to higher angles, moving the positions of the monopoles at the top and bottom of the vortex projected onto the image plane further apart to reduce its effect.

It will also be noted that the concept of a critical tilting angle was entirely circumvented in the more sophisticated treatment used to derive the phase shift in section 6. This is because, effectively, the vortex is split into a bundle of infinitesimal flux lines and the phase shift calculated by counting the number of lines crossed. Whether the number of flux lines crossed was limited by the specimen thickness or the vortex width was automatically accounted for.

### 7.2 Fraunhofer Diffraction Patterns from Flux Vortices

Having calculated an approximate figure for the scattering angles involved in diffraction from flux vortices, we now simulate the expected diffraction pattern. Once the phase shift due to a vortex $\phi(x,y)$ has been calculated (see section 6), it is a simple matter to calculate the exit wavefunction (taking the intensity of the wave to be unity).

$$\psi(x, y) = e^{i\phi(x,y)} \tag{57}$$

The Fraunhofer diffraction pattern is then the squared modulus of the Fourier transform of the wavefunction.

$$I(\mathbf{k}) = |\Psi(\mathbf{k})|^2 \tag{58}$$

This is straightforward to compute and the only problem is that the phase from a vortex cannot be made the same at the boundaries of the image and artefacts occur due to the discrete Fourier transform. To avoid these artefacts, the wavefunction was apodised using a Hanning window before being Fourier transformed. In other simulations we perform, this boundary problem can be allowed for by basing simulations on a pair of oppositely polarised vortices (see Figure 24(a)) but for the diffraction patterns, it is then difficult to see which features arise from which vortex and it is necessary to use a single vortex. Other authors have removed a uniform phase ramp prior to Fourier transforming to reduce the effects of the phase jump at the boundaries [55]. As discussed below, this shifts the origin in Fourier space but for simulations of single vortices, its effect is very minor provided that the size of the image is much larger (30-40 times) than the penetration depth [55].

Figure 19 shows diffraction patterns simulated at different specimen tilts for niobium and BSCCO. The diffraction pattern is asymmetric about the origin meaning that more electrons get scattered in one direction than another as expected for a magnetic object with one prevailing field direction and it can be seen that the estimates of the deflection angle were reasonable. The greater the angle of specimen tilt, the more deflection is observed so to observe the most dramatic effects, the specimen must be tilted to an angle as high as practicable.

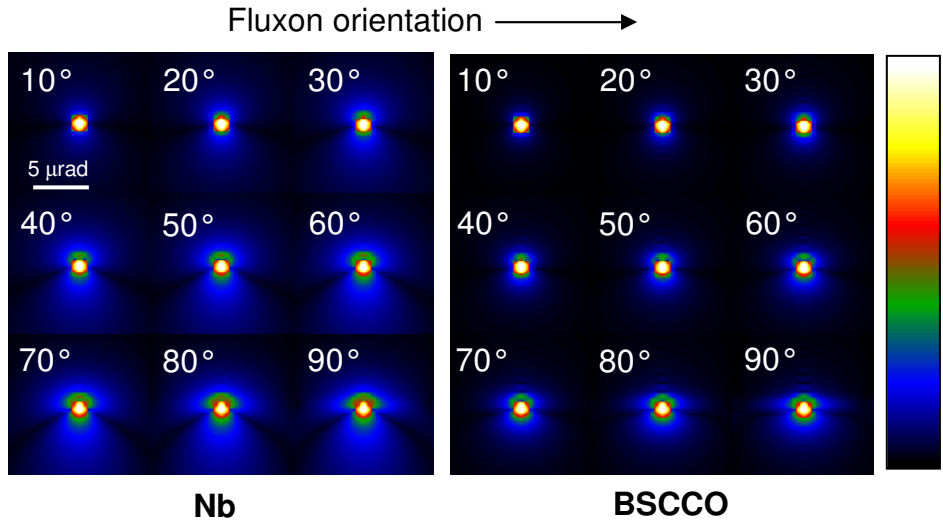

**Figure 19:** Simulated diffraction patterns for single vortices in 200 nm thick Nb and BSCCO displayed as the 4th root of the intensity in false colour according to the linear colour-scale shown on the right. The angles refer to the specimen tilt, $\alpha$.

Figure 19 demonstrates that the diffraction pattern is quite sensitive to the nature of the vortex as can be seen by comparing the diffraction patterns of Nb and BSCCO. In order to observe these patterns experimentally, the camera length would need to be ~500 m. Beam divergence is a significant limiting factor in the observation of the diffraction patterns, effectively convolving the diffraction pattern with a disc whose radius is the beam divergence angle. It can be seen that to distinguish the patterns here, the beam divergence would need to be less than ~0.5 µrad. We have demonstrated that it is possible to realise these conditions experimentally as discussed in section 8.2.

We now generalise these results for a single vortex to a lattice of vortices. To simulate an array of vortices, the phase $\phi_1(\mathbf{r}-\mathbf{R}_i)$ from each vortex centred at position $\mathbf{R}_i$ should be added, giving

$$\phi(\mathbf{r}) = \sum_{\mathbf{R}_i} \phi_1(\mathbf{r}-\mathbf{R}_i) \tag{59}$$

so that the exit wavefunction becomes

$$\psi(\mathbf{r}) = \prod_{\mathbf{R}_i} \exp\left(i\phi_1(\mathbf{r} - \mathbf{R}_i)\right) \tag{60}$$

It is not immediately obvious how to proceed from here to find how the diffraction pattern from one vortex relates to the diffraction pattern of a lattice of vortices and part of the problem is that the phase shift from a vortex lattice is non-periodic. Figure 20 shows the phase from an array of vortices calculated simply by adding the phases of single vortices placed at different positions. It should be noted that the contribution from vortices outside the picture have been added as the B-fields from vortices are long-ranged as discussed in ref. 56.

Although it is not possible to make the phase shift from a single vortex match at the boundaries by subtracting a uniform phase ramp, it is possible to make the phase from a lattice of vortices periodic. This can be seen with reference to Figure 20. If we consider the neighbouring lattice points A, B, C and D, it can be seen that the phase profile along a straight line trajectory from A to D will be the same as the phase profile from B to C as the trajectories pass through identical physical environments although the starting phases may be different in general. The same argument can be made for the trajectories AB and DC. If a phase ramp that matches the phases at the corners of the parallelogram ABCD is subtracted, each lattice point will have the same phase and the phase profiles along parallel edges will be the same. Thus the phase for a lattice of vortices can be made periodic by subtracting a phase ramp.

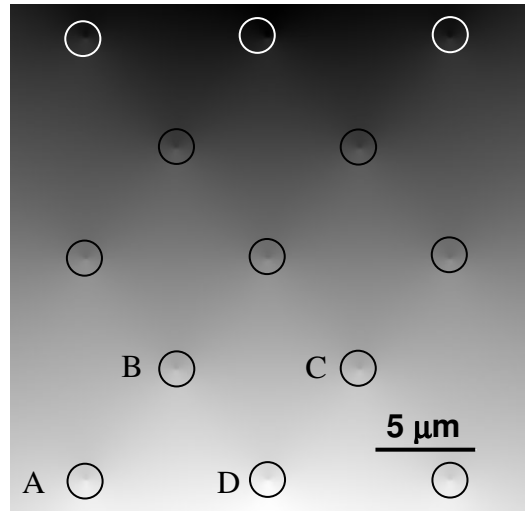

**Figure 20:** Simulated phase profile from a hexagonal lattice of vortices in 200 nm thick Nb tilted to 45° with respect to the electron beam. Each vortex is circled. The lattice parameter is 9.1 µm and note that the lattice has been compressed as it is the projection from a tilted specimen. See text for details.

Physically, the phase ramp is caused by the average B-field associated with the vortex lattice and after its removal, the phase shift can be split into repeating unit cells and we denote the phase within each cell $\phi_{\text{rep}}(\mathbf{r})$. This function is only non-zero inside the unit cell. Denoting the gradient of the phase ramp $2\pi\mathbf{q}$, the total phase shift from the vortex lattice can be written as a phase ramp plus a periodic part:

$$\phi(\mathbf{r}) = 2\pi \mathbf{q}.\mathbf{r} + \sum_i \phi_{\text{rep}}(\mathbf{r} - \mathbf{R}_i) \tag{61}$$

The electron wavefunction can then be written in terms of the wavefunction associated with each unit cell, $\psi_{\text{rep}}(\mathbf{r})$, as:

$$\psi(\mathbf{r}) = e^{2\pi i \mathbf{q}.\mathbf{r}} \left( \sum_i \psi_{\text{rep}}(\mathbf{r} - \mathbf{R}_i) \right) = e^{2\pi i \mathbf{q}.\mathbf{r}} \psi_{\text{rep}}(\mathbf{r}) * \left( \sum_i \delta(\mathbf{r} - \mathbf{R}_i) \right) \tag{62}$$

and by using the convolution theorem, we can give the wavefunction in reciprocal space as:

$$\Psi(\mathbf{k}) = \delta(\mathbf{k} - \mathbf{q}) * \left( \Psi_{\text{rep}}(\mathbf{k}) \left( \sum_i \delta(\mathbf{k} - \mathbf{G}_i) \right) \right) \tag{63}$$

where $\mathbf{G}_i$ are reciprocal lattice vectors.

We can now appreciate the salient features of the diffraction pattern. First, the delta function at the start moves the origin of the pattern to $\mathbf{q}$ which describes the gradient of the phase ramp. The Fourier transform of the wavefunction for a single repeat of the array, $\Psi_{\text{rep}}(\mathbf{k})$, is then multiplied by an array of delta functions so that it is sampled at reciprocal lattice vectors $\mathbf{G}_i$ and, as one would expect, the diffraction pattern for a lattice of vortices appears as regularly spaced spots. The intensity of each spot is determined by the squared modulus of $\Psi_{\text{rep}}(\mathbf{k})$.

It is tempting to equate $\Psi_{\text{rep}}(\mathbf{k})$ to $\Psi_1(\mathbf{k})$, the wavefunction for a single, isolated vortex with an appropriate ramp removed and to say that the diffraction pattern from a lattice of vortices is the same as the diffraction pattern for a single vortex but discretely sampled and with its origin shifted as was done in an early paper [57]. In fact, this is an approximation valid only for widely separated vortices. Whereas $\Psi_1(\mathbf{k})$ extends to infinity, $\Psi_{\text{rep}}(\mathbf{k})$ is zero outside each unit cell. In the isolated vortex case, it is only far from the core that the boundary phase can be made close to zero by subtracting a ramp [55]: as the density of vortices increases, this approximation breaks down.

Figure 21 shows simulated diffraction patterns from hexagonal vortex lattices beginning with a single vortex in (a) and the fluxon density doubling in successive panels thereafter and the expected trends we describe above are seen. At low densities, the basic structure of the diffraction pattern does not change greatly but becomes sampled at discrete points. At the same time, the pattern is shifted downwards owing to the average B-field from the array. As the density of fluxons increases and they overlap more, the shape of the modulating function changes and information about the phase shift from a single vortex is lost This is not a quirk of the diffraction pattern but applies to all images of fluxons: once the density of fluxons is too great (i.e. when their spacing becomes comparable to their penetration depth), information about the nature of single vortices is lost. Throughout the remainder of this section we simulate only single vortices and the results will apply to fluxons spaced much further apart than their penetration depth.

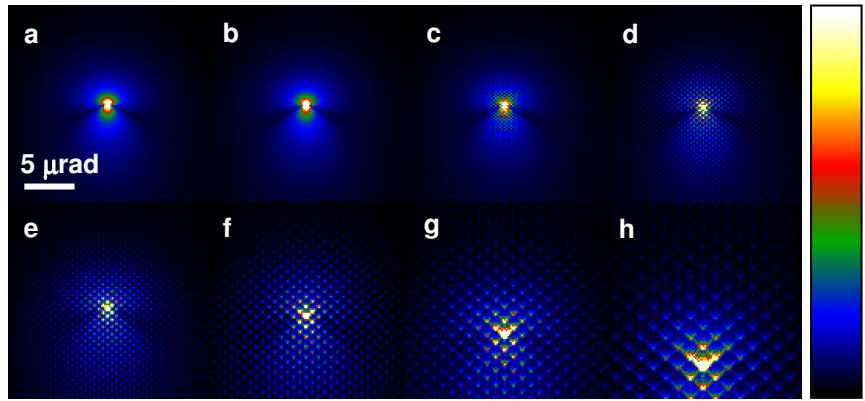

**Figure 21:** Simulated diffraction patterns from a hexagonal array of vortices in niobium tilted at 45° to the beam direction displayed as the 4th root of the intensity in false colour according to the linear colour-scale shown on the right. (Note that the lattice is compressed in the vertical direction as the sample is tilted.) (a) single vortex, (b) 12.8 μm lattice parameter, (c) 9.1 μm, (d) 6.4 μm, (e) 4.5 μm, (f) 3.2 μm, (g) 2.3 μm (h) 1.6 μm.

The simulations shown here were performed by the simple, if inelegant, method of generating the phase for an array of vortices by adding the phases from single vortices at different positions, including vortices outside the image; calculating the wavefunction; apodising with a Hanning window; Fourier transforming and taking the squared modulus. A more elegant method is described by Beleggia and Pozzi [56] where the phase shift is calculated analytically in Fourier space so that the effect of vortices outside the image is accounted for automatically. This approach runs into difficulties at the origin where the Fourier transform is singular but by comparing simulations performed in real and reciprocal space, the authors show that if the point at the origin is set to zero, this is equivalent to subtracting a phase ramp. The authors describe the ramp as 'unessential' but it is caused by the average B-field of the vortex lattice and has the effect of moving the origin of the diffraction pattern. This has the physical consequence that if one were to look at a diffraction pattern from a vacuum and then move to a region of specimen containing an array of vortices, the central beam would shift. It also means that when an image is formed, the electron beam passes through the lens at an angle rather than down the optic axis. With reference to the defocusing phase plates shown in Figure 22 (b) and (c), the brightest diffraction pattern would not pass through the centre of the phase plate and neglecting the phase ramp would change the phase shift that the lens applies to each beam. If the Fourier space approach is used, the effects of the phase ramp can be allowed for either by shifting the diffraction pattern or adding a ramp to the phase in real space.

We conclude that diffraction patterns are sensitive to the detailed structure of individual vortices and this could be a good way to examine their inner structure. However, there is no direct method for working back from the diffraction pattern to the vortex structure and obtaining it will require an iterative process of first guessing a structure, simulating a diffraction pattern from it, comparing it to the experimental result and refining the guess. Simulations have shown that with a regular array of vortices, the diffraction pattern broadly resembles that for a single vortex but discretely sampled. Thus to deduce the structure of a single vortex from a diffraction pattern containing several vortices, it is necessary to look at a sparse lattice or, better, a sparse disordered array of vortices. Obviously, unlike the imaging techniques to be discussed next, the diffraction pattern provides little information on the dynamics of the vortices.

It should be noted that diffraction patterns from vortex lattices in Nb have been recorded by Yoshida *et al.* [20, 57]. These were taken at high fields (for these experiments) of 200–450 G and consist of 6 reflections surrounding the central beam. This gives information on the arrangement of the vortices in the lattice but ideally one would want to take diffraction patterns from much less dense lattices to give information on single vortices.

### 7.3  Fresnel Imaging

Fresnel or out-of-focus imaging is the most widely used method for imaging vortices and has been successfully employed in this investigation to image vortices in BSCCO as described in sections 8 and 9. Defocusing an image is identical to placing a phase plate in the diffraction plane of the microscope which multiplies the wavefunction in Fourier space by $\exp[i\chi(k)]$ where $\chi(k) = \pi \Delta f \lambda k^2 + \frac{1}{2}\pi C_S \lambda^3 k^4$ and $\Delta f$ is the defocus, $\lambda$ the electron wavelength and $C_S$ the coefficient of spherical aberration. Note that even though the Lorentz lens has $C_S = 8$ m, the aberration term is negligible compared with the defocusing term at the defoci used to image flux vortices.

Very large defoci (1-50 cm) are used to image vortices and Figure 22 illustrates why this is necessary. It shows both the simulated diffraction pattern and the defocusing phase plate. It can be seen that for defoci smaller than 1 cm, the phase plate is almost flat in the region where the diffraction pattern has significant intensity and this would give a featureless image. It is only for defoci larger than 1 cm that the phase plate changes significantly over the range of wavevectors for which the diffraction pattern has significant intensity.

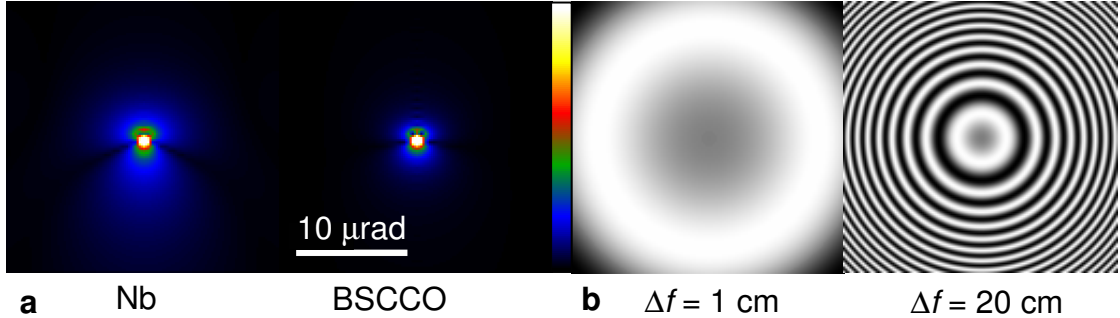

**Figure 22:** (a) Diffraction patterns from Nb and BSCCO displayed as the 4th root of the intensity in false colour according to the linear colour-scale shown on the right. (b) The defocusing phase plate, shown at defoci of 1 and 20 cm. It is contoured by plotting $\sin(\chi(k))$ so that moving from the centre of a black ring to an adjacent white ring represents a phase shift of $\pi$.

Prior to describing the simulations, we discuss a highly simplified model by Harada *et al.* [58] from which we can explain some of the features of Fresnel images of flux vortices. This model is similar to the ray diagram frequently drawn to illustrate the contrast mechanism for Fresnel images of ferromagnetic domains [59]. Figure 23 shows electrons passing through a specimen containing a flux vortex and those which pass near to the vortex are deflected by its associated B-field. In this model only electrons which pass within a radius $W$ of the vortex are deflected. In a defocused image, this creates a black region beneath the vortex where there is a deficiency of electrons and a white region where there is an excess of electrons. The schematic linescans to the right of the figure show the electron excess and deficiency. At zero defocus, the excess electrons coincide with the electron deficiency and so no contrast is observed but as the defocus is increased the electron excess and deficiency move apart as illustrated and the vortex appears as a black-white feature.

It is also possible to see that whereas the image contrast (defined as the difference between the maximum and minimum intensities in the image divided by their sum) improves with increasing defocus, the improvement diminishes as the defocus increases until there is little benefit in increasing the defocus as the improvement is so small. As illustrated by the schematic linescans to the right of the figure, the contrast will improve steadily with increasing defocus until the peak of the electron excess is well separated from the trough of the electron deficiency. In a similar manner to the Rayleigh criterion, we define this defocus as the point at which the separation of the excess and deficiency is equal to their width, $2W$. Beyond this defocus, the black and white features get further apart but the contrast increases very slowly as the maximum of the electron excess is added to the small values in the tail of the electron deficiency. We describe this as the optimal defocus as increasing the defocus beyond this value gives only small improvements in the contrast but makes the spatial resolution (which can be taken as the separation of the black and white features-see the end of this section) worse.

We can estimate this optimal defocus using the characteristic deflection angle $\Theta$ from equations (55) and (56). Since the scattering angle is much less than 1 radian, the separation between the black and white features is

$$\Delta x = \Theta \Delta f \qquad (64)$$

The optimal defocus occurs when $\Delta x = 2W$ giving

$$\Delta f_{\text{optimal}} = 2W / \Theta \qquad (65)$$

Using the scattering angles calculated in section 7.1, this formula gives the optimal defocus for Nb as 7 cm and 21 cm for BSCCO for a 200 nm thick specimen tilted to $\alpha = 50°$.

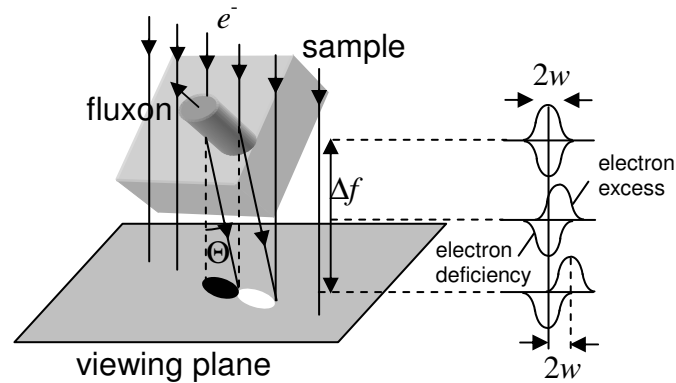

**Figure 23:** Simple model to explain the appearance of Fresnel images. See text for details.

We now discuss simulated Fresnel images of individual flux vortices. The appearance of Fresnel images can be found by calculating the wavefunction from the phase shift calculated in section 6.2, Fourier transforming the wavefunction and multiplying by the phase plate $\exp[i\chi(k)]$ and then inverse transforming. To avoid artefacts associated with the non-periodic nature of the phase shift associated with the vortices, we began with the phase shift from two oppositely polarised vortices as shown in Figure 24(a). The phase from each vortex was not superposed, the phase shift from the top half referred to the top vortex and the phase from the bottom half referred to the lower vortex.

In keeping with experimental results, Figure 24(b) and (c) show that defocused images of vortices appear as black-white features. Figure 24(d) and (e) show the contrast in the image as a function of defocus and it can be seen that in keeping with the predictions of the simple model, the contrast increases with defocus but the improvement in contrast rapidly diminishes beyond a certain point. There is no single point after which it is pointless to increase the defocus further but it can be seen that the earlier estimates of an optimal defocus are reasonable. It can also be seen that the contrast increases as the specimen is tilted and experimentally, the sample should be tilted to as high an angle as is practicable. Similar graphs are shown in ref. 60 and the effect of B-field spreading near the sample surfaces discussed in section 6.3 is also included.

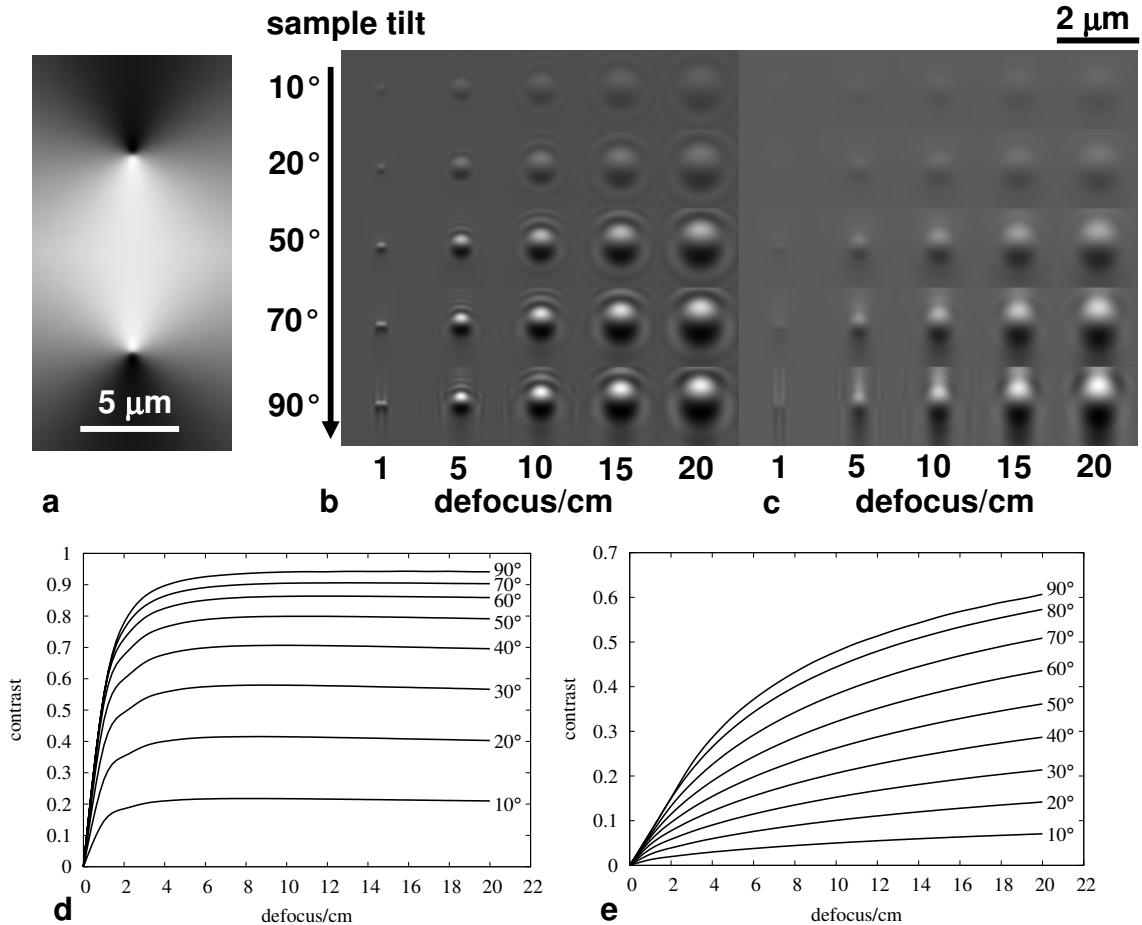

**Figure 24:** (a) The phase shift from a vortex-antivortex pair used as a basis for these simulations to make the phase shift periodic. Note that the phase shifts are not superposed so the top half of the figure refers to the upper vortex and the lower half to the lower anti-vortex. (b) the appearance of Fresnel images for vortices in Nb. (c) the same for BSCCO. (d) The contrast of vortex images as a function of defocus for Nb and (e) for BSCCO.

The spatial resolution of the technique is rather hard to judge as the contrast of the image, not just its visual appearance, contains information on the width of the vortex but extracting this requires prior knowledge of the B-field in the vortex. If we use only the visual appearance of the image, the situation is as illustrated in Figure 25(a) where simulations of Fresnel images are shown for Nb, BSCCO and a vortex of infinitesimal radius and each image has been scaled to have the same maximum and minimum. This shows that at defoci above 1 cm, the images appear very similar although the penetration depths are very different. Since a vortex of infinitesimal radius still produces a Fresnel image with a finite width, the spatial resolution of the technique can be taken as the width of the black-white feature normal to the vortex direction.

There is no unique way to quantify the width of the black-white feature and we use a quantity analogous to the full-width-half-maximum. More precisely, we take a line trace through the centre of the black-white feature normal to the vortex and subtract the mean value. One extremity of the image is taken as the point at which the value falls to half its maximum and the other extremity is the point at which the value is half the minimum (which is negative, having subtracted the mean) and the width is the distance between these extremities. Figure 25(b) shows how this apparent width for an infinitesimal vortex varies as a function of defocus and this can be taken as a measure of the resolution of the technique. At 1 cm, about the smallest defocus at which vortices have been observed in the literature, the width of the infinitesimal vortex is about 200 nm and bearing in mind the qualifications mentioned earlier, we take this value as a measure of the best resolution of the technique.

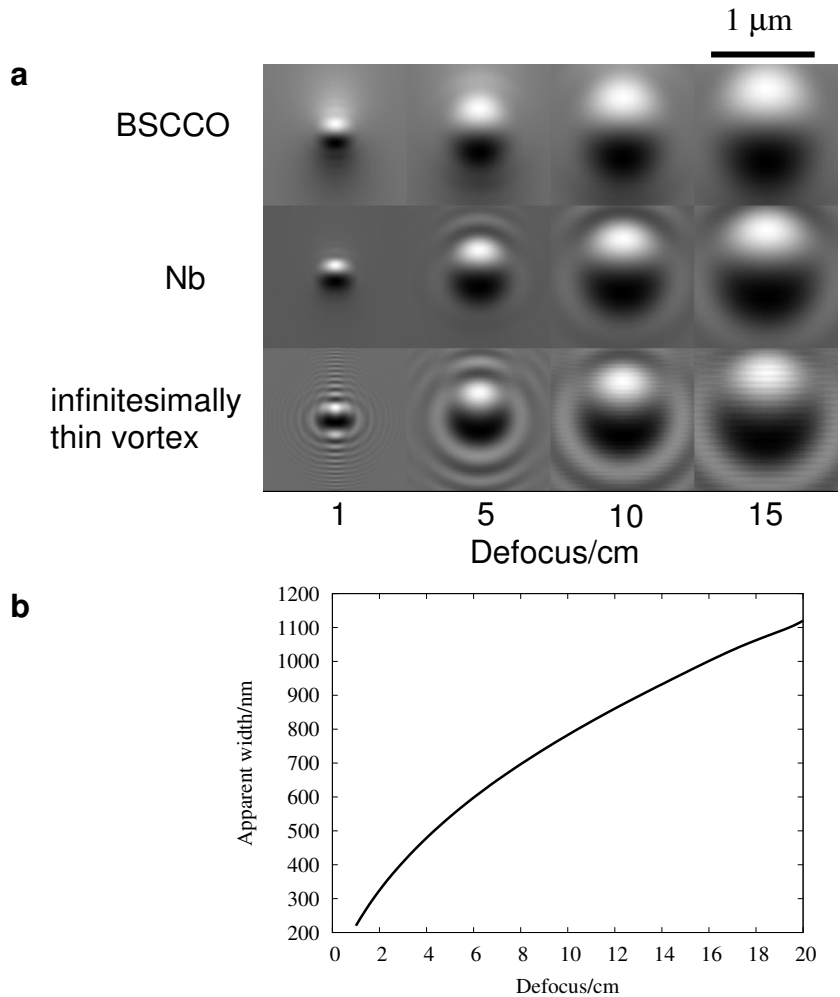

**Figure 25:** (a) Fresnel Images for BSCCO, Nb and a vortex of infinitesimal radius at 50° sample tilt. Each image has been scaled to have the same maximum and minimum. (b) The apparent width of Fresnel images of infinitesimal vortices as a function of defocus.

It can be seen then that Fresnel images are not very sensitive to the structure of vortices unless the changes in the structure are dramatic. Two examples from the literature of changes in vortex structure which are sufficiently drastic to be observed in Fresnel images are vortices which align along columnar defects [61] and vortices in YBCO which apparently abruptly change from a tilted stack of pancake vortices to a sequence of pancake and Josephson vortices [43].

We now assess the minimum B-field to which Fresnel imaging is sensitive. The contrast in a given image is determined by the deflection angle (equation (52)) which is proportional to the product of the B-field normal to the electron beam and the sample thickness parallel to the electron beam (called the B-field-thickness product), $B_\perp t = B_\perp l / \cos\alpha$. The contrast in a Fresnel image can always be improved by moving further out of focus and with our microscope, we have found that 30 cm is about the largest defocus we can have before the image becomes severely distorted. In accordance with Beleggia [45], we took the minimum detectable contrast to be 0.03 and carried out simulations to find the widest vortex which was still above this minimum contrast at 30 cm defocus. We found that in a 200 nm thick specimen tilted to 50°, a vortex with a penetration depth of 2500 nm gave a contrast of 0.03 at 30 cm defocus. Using equation (52), this has a B-field normal to the electron beam of 0.0003 T (3 G) so the minimum detectable B-field thickness product is 0.09 Tnm.

*7.4    Foucault Imaging*

An alternative way to visualise the magnetic structure of a specimen is Foucault imaging where a beam stop (usually the edge of the objective aperture) is inserted in the back focal plane, blocking some of the deflected rays. Figure 26 shows a schematic of its operation imaging two magnetic domains.

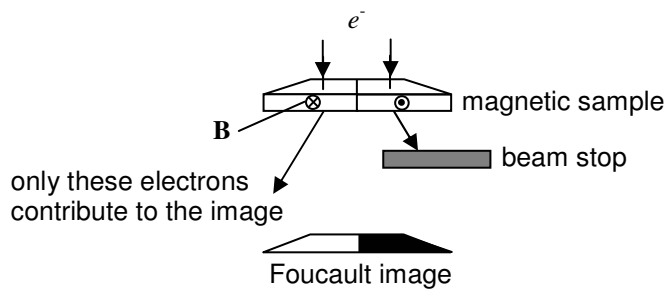

**Figure 26:** Schematic illustration of Foucault imaging from 2 magnetic domains. The domains deflect the electron beam in different directions and one of these is blocked by a beam stop. Only electrons deflected in one direction contribute to the image so that one domain appears dark.

The example in Figure 26 is slightly misleading in that it ignores interference effects which are observed with a coherent electron beam. Johnston and Chapman [62] have demonstrated that when Foucault images of ferromagnetic specimens are taken under coherent illumination, the bright parts of the images resemble the cosine of the phase images often used in holography to visualise magnetic fields although the two are not mathematically equivalent.

To simulate Foucault images of flux vortices, we Fourier transform the exit wavefunction to give the wavefunction at the diffraction plane, set pixels equal to zero to represent the effect of a straight-sided semi-infinite beam-stop then inverse transform and take the squared modulus to give the Foucault image. As with the simulations of Fresnel images, we begin with the exit-wavefunction from a vortex-antivortex pair as this is a periodic object as required for Fourier transforming. The results are shown in Figure 27 and it can be seen that the appearance of the image varies greatly as a function of the beam-stop displacement. Similar simulations of Foucault and phase plate images (discussed in section 7.5) are described in ref. 20.

To assess the visibility of vortices imaged with coherent Foucault imaging we plot a quantity we term the signal-contrast against the aperture position in Figure 28. The reason for this new quantity is that Foucault imaging uses an aperture to block the electrons and so the number of electrons reaching the viewing screen depends on the aperture position. The simulations are performed so that a bright field image (which would be featureless) would have an intensity of 1 unit per pixel. The signal-contrast is the average value of each pixel (representing the proportion of electrons reaching the screen) multiplied by the contrast. These plots are directly comparable with the contrast versus defocus plots for the Fresnel images where the signal is always 1.

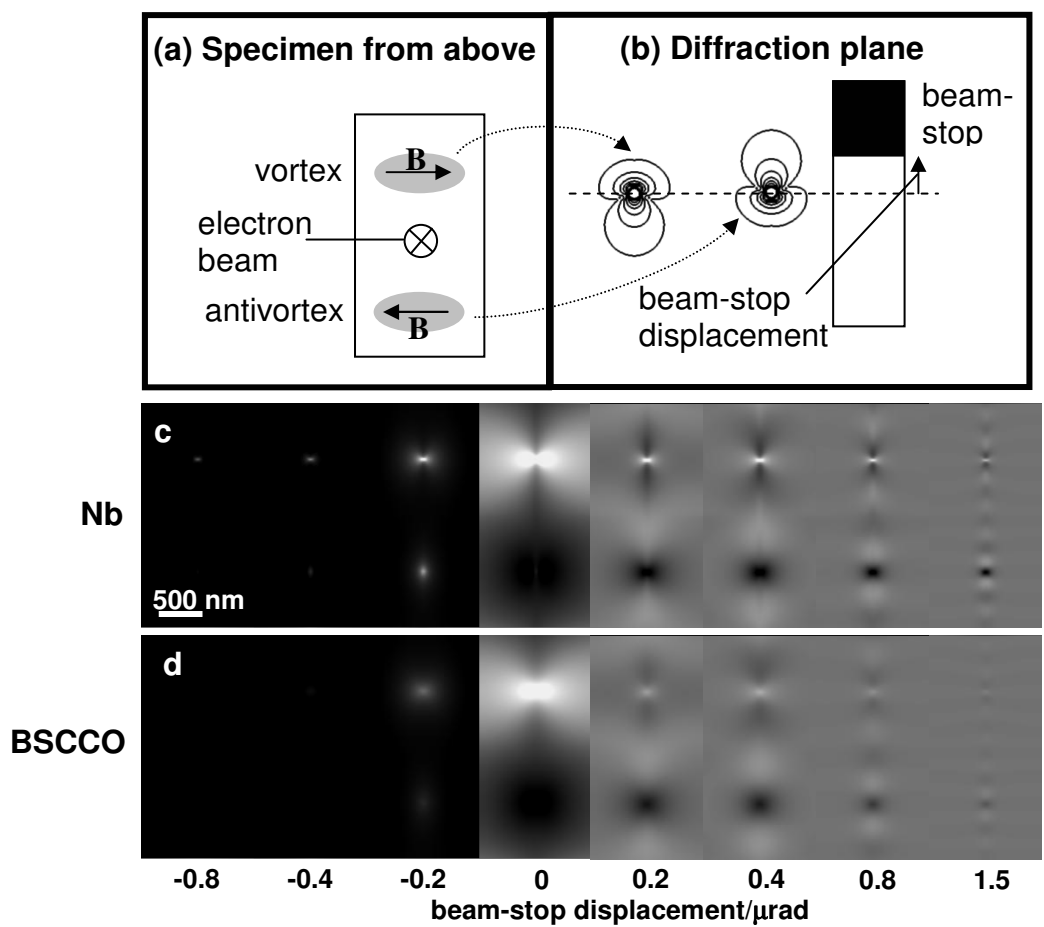

**Figure 27:** Simulated Foucault images of vortices. (a) Simulations were carried out for a vortex-antivortex pair as this gives the periodic boundary conditions needed for Fourier transforming (see Figure 24(a)). (b) In the diffraction plane, each vortex produces a diffraction pattern shown by the contour lines. The simulations were carried out for different aperture displacements measured from the origin with positive values indicating that the open part of the aperture is beyond the origin (indicated by the dashed line). (c) the appearance of simulated images for different aperture displacements for 200 nm thick Nb and (d) BSCCO tilted to 50°. Note that the intensity scaling is the same in both.

It can be seen from Figure 28 that, provided the aperture is placed close (within 4 μrad) to the most intense part of the diffraction pattern, images with a significantly higher signal-contrast compared with Fresnel images are possible. An even higher contrast is achievable using a phase plate as discussed in the next section.

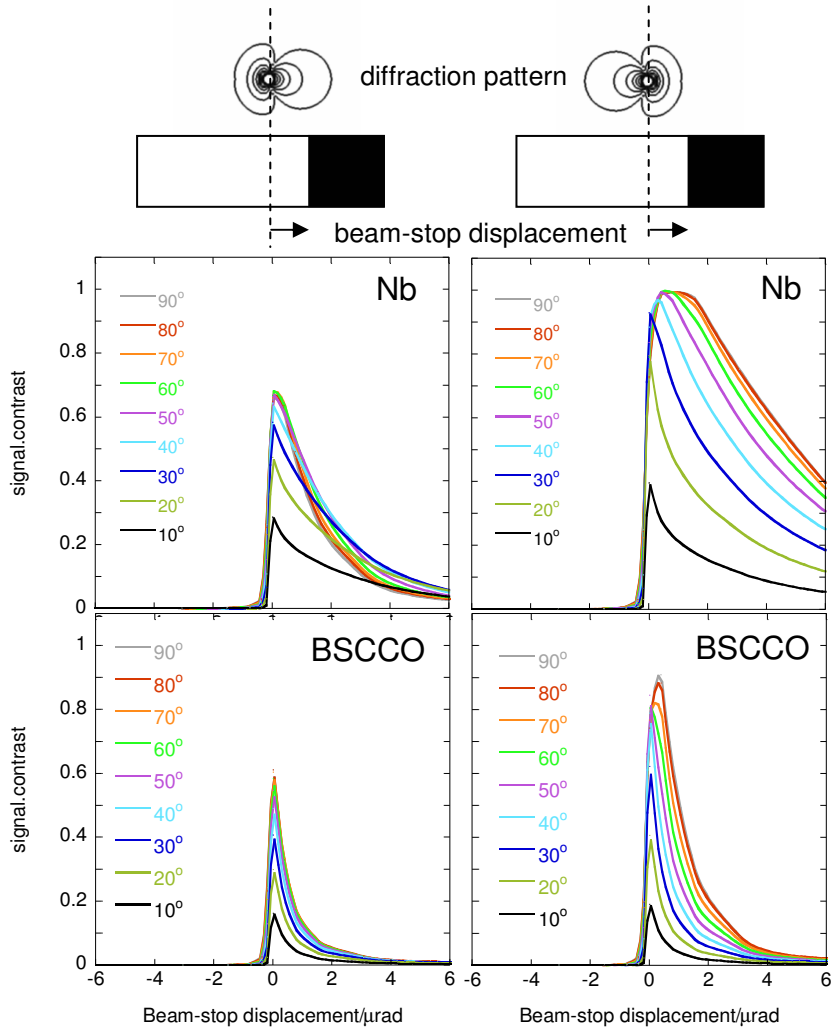

**Figure 28:** Signal-contrast for Nb and BSCCO as a function of aperture displacement.

*7.5    Phase Plate Imaging*

Phase plate imaging is very similar to Foucault imaging but instead of absorbing electrons with a beam-stop, a plate is placed in the diffraction plane which phase-shifts some of the electrons. The phase plate is usually an amorphous carbon film of an appropriate thickness to produce the required phase shift but ideally leave the amplitude of the wave unaltered.

Phase plates are used in transmission electron microscopy to enhance the contrast from biological specimens [63] and have also been used occasionally to visualise B-fields in ferromagnetic samples [62]. The two main types of phase plate which have been used are the Zernike and Hilbert geometries. The Zernike phase plate consists of a small hole in the carbon film which lets through only the central beam of the diffraction pattern and phase shifts the rest of the diffraction pattern. The Hilbert phase plate is a straight edged plane which phase shifts a rectangular potion of the diffraction pattern and leaves the rest unchanged.

The smallest objective aperture currently fitted to the CM300 electron microscope used here has a physical diameter of 5 µm and its size in reciprocal space is approximately 130 µrad. To image vortices, the Zernike phase plate would need to be ~1 µrad and thus have a physical size of ~40 nm using the same optical arrangement as we have. It seems that it would be difficult to manufacture a Zernike phase plate suitable for imaging vortices without redesigning the microscope optics. The Hilbert plate on the other hand has the same geometry as the beam-stop used in Foucault imaging and as Foucault imaging of vortices has already been demonstrated [20], it should be possible to use a Hilbert phase plate to image vortices.

Figure 29 shows simulated images from a $\pi$ shifting Hilbert phase plate. It can be seen that the appearance of the images is similar to that of the Foucault images but the signal-contrast is higher for a broader range of displacements of the plate from the origin of the diffraction pattern. It should be noted that in weak phase objects such as biological specimens, the optimal phase plate produces a phase shift of $\pi/2$. Flux vortices are not weak phase objects, however, and as highlighted in a recent paper by Beleggia [64], the phase shift that gives maximum contrast in other objects depends on the object under investigation. We have performed simulations throughout the whole range of phase shifts and found that the $\pi$ shifting plate gives the best contrast for Nb and BSCCO for all specimen tilts.

Figure 30(a) shows the results for 200 nm thick Nb and BSCCO specimens tilted to 50° which provides a useful contour map showing how accurately the phase plate needs to be placed and how close to $\pi$ the phase shift needs to be to maximise the contrast. To achieve a contrast of 0.8 or better in 200 nm thick Nb, the phase plate must be within about 2 µrad of the origin and the phase shift should be within 1 radian of $\pi$. To achieve the same in BSCCO, the phase plate must be within 0.5 µrad of the origin and produce a phase shift within 1 radian of $\pi$. As shown in Figure 30(b) and(c), the signal-contrast is better than that which can be achieved by Fresnel and Foucault imaging (see sections 7.3 and 7.4) and it is interesting to note that they are quite insensitive to the phase shift produced by the phase plate.

Given that the mean inner potential of carbon is 6.8 V [65] and that good contrast can be achieved even if the phase shift is 1 radian away from $\pi$, the optimal phase plate should have a thickness of 71 nm but can be in error by ±24 nm and still achieve good contrast. It is more likely that poor image contrast will result from the difficulties in positioning the aperture.

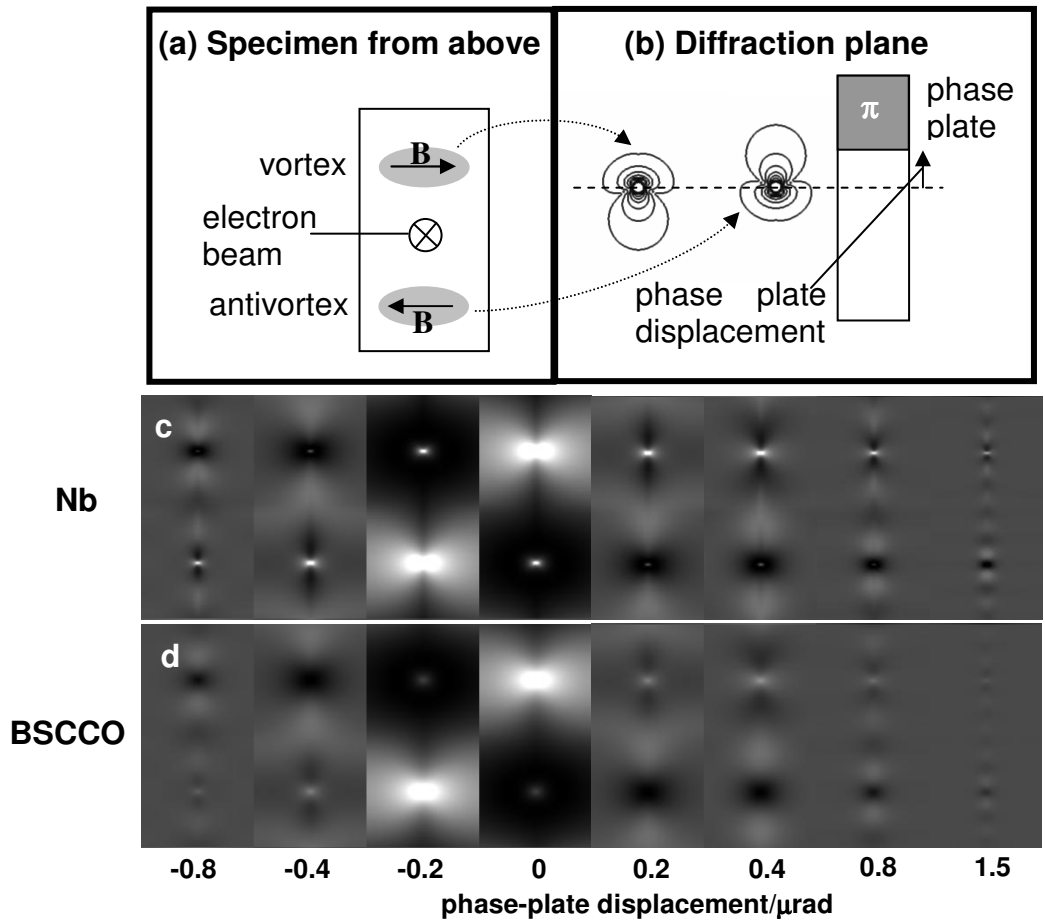

**Figure 29:** Simulated images of vortices taken using a π-shifting Hilbert phase plate. (a) As described before, simulations were carried out for a vortex-antivortex pair. (b) In the diffraction plane, each vortex produces a diffraction pattern shown by the contour lines. The simulations were carried out for different aperture displacements measured from the origin with positive values indicating that the open part of the phase plate is beyond the origin (indicated by the dashed line). (c) the appearance of simulated images for different aperture displacements for 200 nm thick Nb and (d) BSCCO tilted to 50°. Note that the intensity scaling is the same in both.

In principle, the spatial resolution of the image should be unaffected by the insertion of a Hilbert phase plate as it does not restrict the spatial frequencies present in the image, it simply adds an additional phase shift. The spatial resolution is then limited to the resolution imposed by the spherical aberration coefficient of the lens, 2 nm. In the case of Foucault imaging, all the electrons scattered in one direction are removed from the image but the resolution *per se* is not affected. It should be noted that when the phase plate is accurately positioned, simulations show that the vortex cores of Nb and BSCCO can be distinguished (compare the pictues for -2 μrad and 0 μrad displacement in Figure 29(c), (d)) unlike the Fresnel images of Figure 25.

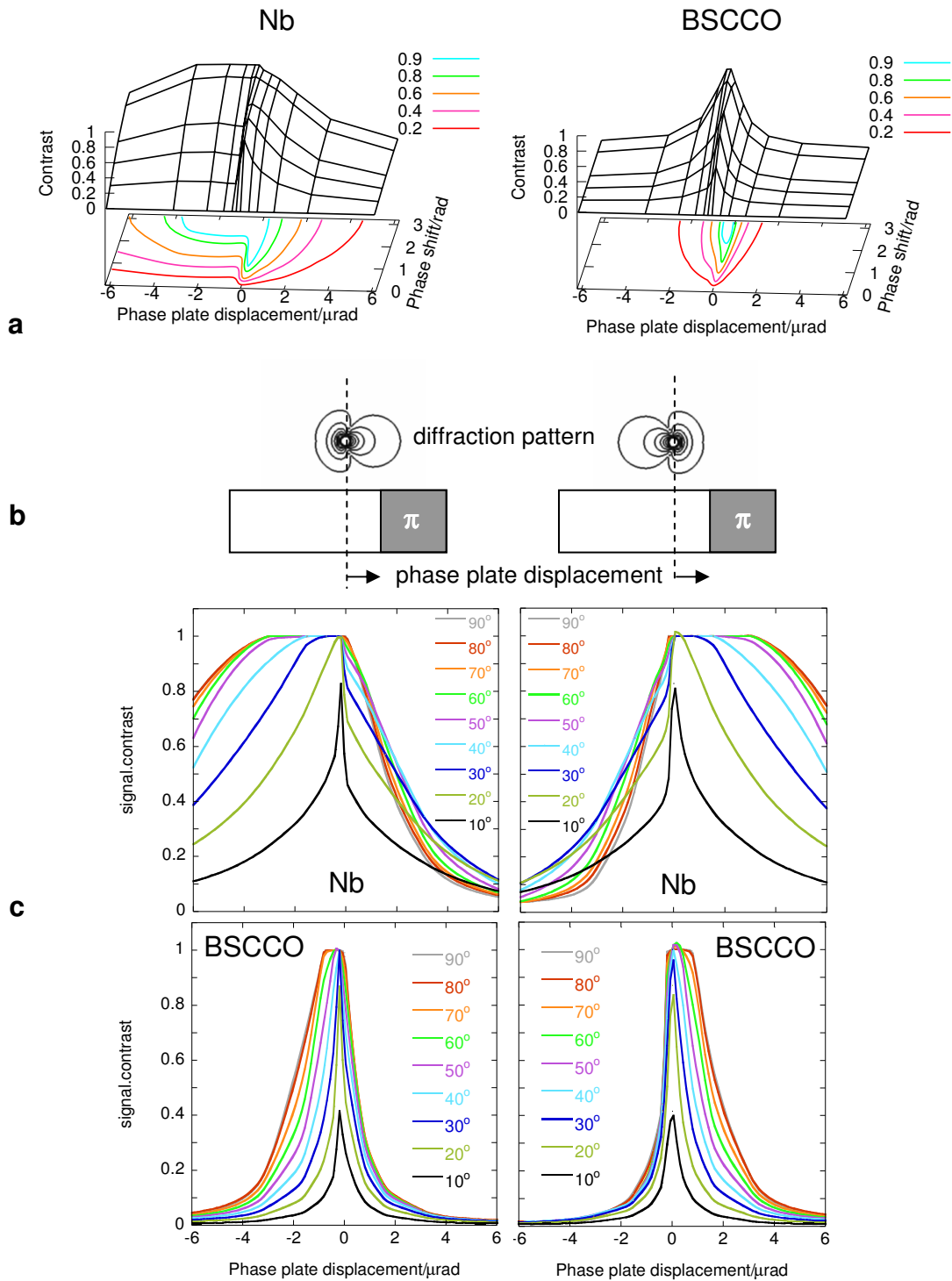

**Figure 30:** (a) 3D map of the contrast for 200 nm thick specimens of Nb and BSCCO tilted to 50° as a function of aperture displacement and phase shift. Contours are shown below. (b) Schematic showing the shape of the diffraction pattern and the positioning of the aperture in the diffraction plane. (c) Plots of contrast versus aperture displacement for 200 nm Nb and BSCCO for a π shifting phase plate for various specimen tilt angles.

The B-field sensitivity can be estimated by considering how much deflection takes place in the diffraction pattern and considering how accurately the phase plate can be positioned. From the diffraction angle in equation (52), the B-field sensitivity is then

$$\delta(B_\perp t) = \frac{h}{e\lambda} \delta\Theta \tag{66}$$

where $\delta\Theta$ is the precision in the aperture positioning. It is hard to judge how accurately the phase plate can be positioned as it depends on the mechanical stability of the aperture arrangement but for a 200 nm thick sample with the aperture misplaced by 0.2 μrad, the B-field sensitivity will be 0.002 T (20 G).

The use of a phase plate looks like a very promising method for imaging vortices as it has a high spatial resolution and B-field sensitivity. It also avoids the need for the large defoci used in Fresnel imaging and produces images with a higher contrast. Phase plate images are not directly interpretable and will require an iterative process of guessing a structure and simulating images to produce quantitative results. Electron holography on the other hand has the ability to produce directly interpretable images.

### 7.6 Electron Holography

All the techniques for imaging vortices are an attempt to record the phase of the electron wavefunction in some form. This is difficult as the process of taking an image records only the intensity and loses the phase information. Unlike the other imaging techniques, the phase can be recovered directly using electron holography.

An electron hologram is recorded by using a positively charged wire, called an electron biprism, to interfere electrons which went through the specimen with electrons which went through vacuum. Fourier analysis can then be used to extract the amplitude and phase of the electron wavefunction [66]. The phase images produced by this technique will look like the phase image in Figure 16(d) and will contain a great deal of information about the structure of individual vortices subject to the resolution constraints which we now consider.

The spatial resolution of the Lorentz lens is ~2 nm but in practice, the resolution of the technique is limited by the signal-to-noise ratio of the hologram which is determined by the visibility of the holographic fringes and the shot noise in the CCD camera. According to Lichte *et al.* [67], this puts a limit on the smallest phase which can be measured between two pixels of

$$\delta\phi = \frac{\text{SNR}}{\mathcal{V}} \sqrt{\frac{2}{n}} \tag{67}$$

Where SNR is the signal to noise ratio of the detector, $\mathcal{V}$ is the visibility of the holographic fringes and $n$ the number of electrons per pixel.

The signal to noise ratio of modern CCD cameras is ~0.9 and the limiting factor is the fringe visibility which is typically 0.2–0.3.
Thus, the smallest phase gradient measurable between two pixels is

$$\delta\left(\frac{d\phi}{dx}\right) = \frac{\text{SNR}}{\mathcal{V}}\sqrt{\frac{2}{n}}\frac{1}{X} \tag{68}$$

where $X$ is the spacing between two pixels as measured in the image.

The phase can be related to the B-field in the sample by the Aharanov-Bohm formula

$$\phi(x) = -\frac{e}{\hbar}\int \mathbf{B}_\perp . d\mathbf{S} \tag{69}$$

where $\mathbf{B}_\perp$ is the component of the B-field normal to the electron beam and $d\mathbf{S}$ is an element of vector area.

If we consider Cartesian coordinates with $z$ in the electron beam direction and take $B_\perp$ to be in the $y$ direction, this can be written as

$$\phi(x) = -\frac{e}{\hbar}\int_0^x \int_{-\infty}^\infty B_\perp(x',z)dz\,dx' \tag{70}$$

The phase gradient is then

$$\frac{d\phi}{dx} = -\frac{e}{\hbar}\int_{-\infty}^\infty B_\perp(x,z)dz \tag{71}$$

Thus the phase gradient is proportional to the integral $\int B_\perp dz$, the B-field-thickness product. It is frequently written as the B-field averaged over the specimen thickness $t$ so that

$$\bar{B}_\perp(x)t = \int_{-\infty}^\infty B_\perp(x,z)dz \tag{72}$$

Thus, the smallest measurable B-field-thickness product is

$$\delta(\bar{B}_\perp t) = \frac{\hbar}{e}\frac{\text{SNR}}{\mathcal{V}}\sqrt{\frac{2}{n}}\frac{1}{X} \tag{73}$$

This figure can be improved using prior knowledge of the object. An obvious way of reducing the noise is to low-pass filter the image: in effect assuming that the phase changes slowly over the size of the pixel. This effectively combines several pixels into one, reducing the noise but also degrading the spatial resolution. If the low pass filter averages over $m \times m$ pixels, the smallest measurable B-field is

$$\delta(\bar{B}_\perp t) = \frac{\hbar}{et} \frac{\text{SNR}}{\mathcal{V}} \sqrt{\frac{2}{n}} \frac{1}{Xm} \tag{74}$$

The value *Xm* can be regarded as the resolution of the image because if the pixel spacing is smaller than the intrinsic resolution of the Lorentz lens (~2 nm), it is advantageous to smooth or rebin the data over a sufficient number of pixels until the size of one pixel is equal to the intrinsic resolution. That way, the B-field sensitivity is optimised simply by averaging over meaningless information. As vortices are large objects (300 nm in the case of BSCCO), it will often be the case that the size of one pixel is larger than the intrinsic resolution of the lens and then the resolution of the image is simply the size of the smallest pixel. Writing the resolution as $\delta x = Xm$, we obtain a relationship reminiscent of Heisenberg's uncertainty relationship.

$$\delta(\bar{B}_\perp t)\delta x = \frac{\hbar}{e} \frac{\text{SNR}}{\mathcal{V}} \sqrt{\frac{2}{n}} \tag{75}$$

For a signal-to-noise ratio (SNR) of 0.9, a visibility of (*V*) of 0.3 and 1000 counts per pixel (*n*),

$$\delta(\bar{B}_\perp t)\delta x = 88.2 \text{ Tnm}^2 \tag{76}$$

Thus if we want a spatial resolution of $\delta x = 2$ nm, the smallest B-field-thickness product we could obtain under these conditions would be $\delta(\bar{B}_\perp t) = 44$ Tnm. For a 200 nm sample thickness, this gives a B-field sensitivity averaged over the sample thickness of $\delta\bar{B}_\perp = 0.2$ T (2000 G). On the other hand, if we took a low resolution image with $\delta x = 50$ nm, $\delta(\bar{B}_\perp t) = 1.8$ Tnm and for $t = 200$ nm, $\delta\bar{B}_\perp = 0.009$ T (90 G).

Holography may be less sensitive to B-fields than phase plate imaging but unlike every other technique, the B-field can be measured quantitatively without requiring extensive simulations. It has the disadvantage that the field of view of the hologram is limited to about 3 μm in standard Lorentz holography mode but it may be possible to improve on this by non-standard lens configurations. The other disadvantage is that the image must be taken on a part of the specimen next to vacuum so that interference can take place with electrons which traverse the vacuum. A possible alternative which we intend to explore is to interfere two different regions of specimen to give the relative phase shift between the two, a technique described in ref. [68].

## 8. Experimental Methods

### 8.1 Specimen Preparation

Preparing specimens for the observation of flux vortices is the most challenging aspect of the technique as the specimens must have a very large thin area (~30 μm) and have areas which are flat enough that the contrast is not dominated by bend contours. The specimen should also be a single crystal so that contrast from grain boundaries and other defects does not dominate the image.

Polycrystalline BSCCO samples were prepared by Dr T. Benseman at the Cavendish Laboratory, University of Cambridge from $Bi_2O_3$, $SrCO_3$, $CaCO_3$ and CuO by repeated grinding, pelleting, and annealing at increasing temperatures up to 865°C to produce samples with a nominal stoichiometry of $Bi_{2.15}Sr_{1.85}CaCu_2O_{8+\delta}$. The process was repeated until no further impurity phases could be seen in x-ray powder diffraction patterns. These polycrystalline samples were then pressed into rods (5 mm diameter) and single crystals were grown from these using the travelling solvent floating zone method at a rate of 0.2 mm/hour at the Department of Physics, University of Warwick by Dr G. Balakrishnan and Dr T. Benseman. The final single crystals were found to have stoichiometry $Bi_{2.01}Sr_{1.82}CaCu_{2.03}O_{8.2}$ by electron probe microanalysis. The samples were not made specifically for this investigation and are offcuts from samples similar to those described in ref. 69.

BSCCO is a layered material like mica and can be cleaved in a similar way to produce electron transparent samples which are thin in the **c** direction, normal to the superconducting planes. We originally cleaved the samples with sticky-tape (3M Post-It Flags worked best) and although this produced shards with large thin areas, the samples were difficult to mount and residue from the sticky-tape dominated the contrast at high defoci and we were unable to remove this residue either by dissolving it in acetone or plasma cleaning. Our final method was to pull apart shards from the crystal using fine-pointed tweezers. This method was good as no contaminants were introduced but was unreliable as samples with large thin areas were produced only very occasionally but we were unable to find a better method. Figure 31(a) shows the specimen with the largest thin area we produced (~100×300 μm). We found that areas of the specimen which were translucent when observed by using transmission optical microscopy were usually also electron transparent. The samples were stored in a vacuum desiccator as they degrade due to moisture in the air so that after a few days in the air, they cease to be superconducting.

In our electron microscope, it was only possible to tilt the sample holder to about 25° without striking the objective lens pole piece and much higher tilt angles were required in order to have enough contrast to see the vortices. Thus the samples were mounted in such a way that they were already tilted to ~50° when they entered the microscope. This was achieved by cutting and bending 3 mm copper slot grids as shown in Figure 31(b) and (c). Once the sample was loaded into the microscope, we translated and tilted the sample until an area with minimal diffraction contrast was found. Taking a tilt series was impossible as only a very few tilting conditions gave low enough diffraction contrast to see the vortices.

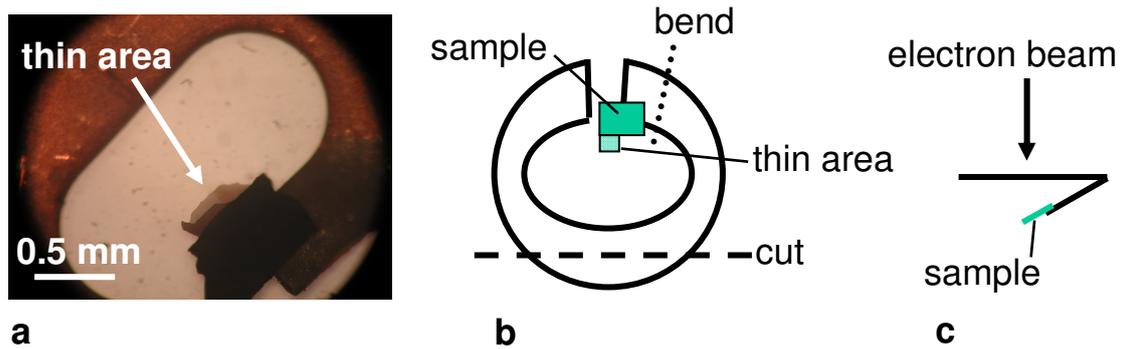

**Figure 31:** BSCCO Sample for Electron Microscopy. (a) Optical micrograph showing BSCCO sample mounted on a copper grid. Both the transmission and reflection lights are illuminated and it can be seen that thin area is optically translucent. (b) Schematic showing how the sample is mounted. The copper grid is originally circular with a hole in it. A slot is first cut from the grid with a razor and the sample glued to one arm with silver glue. The arm is bent to around 50º so that the sample is already tilted when it enters the microscope (shown in cross-section in (c)). The rear part of the grid is also cut away so that the grid fits into the specimen holder.

BSCCO samples are the only sample type that we have attempted so far. The Tonomura group give details on the preparation of niobium by chemical etching in ref. 19 and some details of the preparation of YBCO by focussed ion beam milling in ref. 70.

*8.2    Transmission Electron Microscopy*

In this investigation, microscopy was performed with a Philips CM300 transmission electron microscope. It operates at 300 kV, has a Shottky field emission gun, a Lorentz lens and a Gatan post-column energy filter. Images filtered with a 20 eV energy window were recorded digitally with a CCD camera.

A vertical magnetic field was applied to the sample using the electron lenses. The original intention was to image the specimen with the Lorentz lens and weakly excite the main objective lens to produce a magnetic field. However, this caused the image to vibrate due to what we presume is an electrical fault. We have yet to trace its origin and instead the field was applied by altering the current in the upper twin lens which can be done using 'column-align' mode.

The B-field produced for various lens currents was calibrated by Dr R.E. Dunin-Borkowski by inserting a Hall probe into the specimen entry port. This calibration was performed some years ago and although it was adequate for estimating the lens currents needed to produce a field of a few tens of Gauss, we intend repeating the calibration when more accurate figures are required. The original calibration was performed by venting the microscope column as the Hall probe had no vacuum seal and we are anxious to avoid doing this. A method for converting an old sample holder into a Hall probe is described in ref. 71 and we intend to build our own following these guidelines.

The defocus was calibrated using an Agar Scientific S106 calibration specimen consisting of a cross grating of lines with 2160 lines/mm ruled on an amorphous carbon film. The defocus corresponding to a particular Lorentz lens current was determined by taking the power spectrum of the calibration sample which showed the contrast transfer function as a series of rings as well as spots corresponding to the repetition of the grating. This allowed an absolute measurement of the radius of the rings. Mathematically, the intensity of the rings in the power spectrum is proportional to $\sin^2\left(\pi \Delta f \lambda k^2\right)$ (where $k$ is the coordinate in reciprocal space – note that $C_S$ is negligible at these defoci) and so a measurement of their radii allowed the defocus ($\Delta f$) to be determined. The same specimen was used to calibrate the magnification of the images by noting the lens settings for each image of the superconductor and then taking images of the calibration sample at the same lens settings after the experiment.

It is of interest to note that we also took diffraction patterns from the same calibration grid to ascertain whether it was possible to obtain a high enough camera length and small enough angular resolution to take diffraction patterns from a lattice of vortices or even a single vortex. We achieved this using standard Lorentz mode using a nominal camera length of 60 m. Our Gatan Imaging Filter gave an extra magnification of 25× giving an overall camera length of 1500 m on the CCD camera. It was not necessary to use free lens control mode (where each lens can be operated independently) to achieve this. Whilst this paper was under review, we successfully took a diffraction pattern from a vortex lattice and this will be described in a future publication.

In this experiment we concentrated on taking Fresnel images as this is the simplest technique and the results are reported in section 9. The main problem with the technique is obtaining very large defoci (~20 cm) although with the CM300 electron microscope, this was a matter of reducing the current in the Lorentz lens until a suitably underfocussed image was obtained (it was not possible to apply sufficient current to obtain an overfocussed image with the same defocus). The large underfocus also had the consequence that the image magnification was reduced which was suited for imaging vortices over a large field of view (~20 μm). The defoci used here were in line with the optimal defocus of 20 cm estimated in section 7.3. In the current mode of operation, the image became severely distorted beyond 30 cm defocus.

## 9. Experimental Results

The main result to report in this paper is that electron microscopes have advanced sufficiently that flux vortices can be imaged with an unmodified, commercial transmission electron microscope, opening the field for many other researchers. Here we show Fresnel images of vortices and discuss image processing techniques that are helpful in interpreting the images. We examine a series of images taken as the B-field applied to the sample is reduced and show how correlation analysis is helpful in interpreting the behaviour of the vortices. Similar analyses have been used to extensively examine vortex ordering using the Bitter technique (see e.g. ref. 72). To our knowledge, two such analyses have been performed to date from transmission electron micrographs of niobium, both using the same data [73, 74].

*9.1   Motif Averaging*

We begin by discussing a method for simultaneously identifying the positions of flux vortices in an image and producing an average image of a single vortex, greatly improving the signal-to-noise ratio. This was carried out using the SEMPER image processing package [75] as many of the routines are library programmes but they could be readily coded in other image processing packages.

The technique is shown in Figure 32. The original vortex image (Figure 32(a)) was first Fourier-ring filtered to diminish the effect of large scale contrast variations and small scale noise, leaving the contrast from the vortices (not shown). It was then possible to see the vortices clearly enough to manually extract an image of a single vortex which we term the 'motif'. The image of this single vortex was then cross-correlated with the ring-filtered image resulting in an image whose peaks represent the vortex positions (Figure 32(b)). These peaks can be automatically identified and their positions tabulated using SEMPER's 'peaks' routine which identifies pixels whose value is higher than the surrounding 8 pixels resulting in Figure 32(c). This table of peak positions is then used to extract individual fluxon motifs from the image and then average over all of them. The authenticity of the resulting average fluxon image can be assessed by performing the averaging over every second peak and checking that the final image is not greatly altered. This procedure is performed interactively so that incorrectly identified peaks can be eliminated. We did not interactively add vortex positions so as not to bias the result.

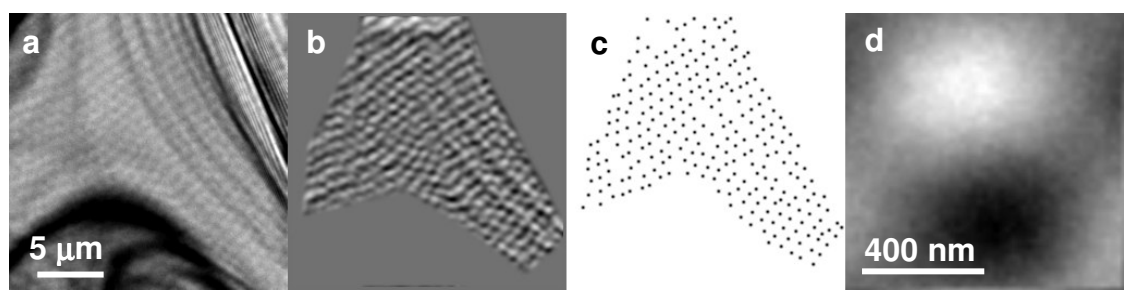

**Figure 32:** Identifying the positions of flux vortices. (a) Unprocessed Lorentz image showing flux vortices in BSCCO taken at 21 K in a field of 28 G at a defocus of 15 cm tilted to 52 ± 2°. A single black-white feature representing one vortex was extracted and then cross-correlated with the original image to give (b). (b) Cross-correlation image where the peaks represent the positions of each vortex. Note that only the middle portion of the image has been used to avoid those areas strongly affected by diffraction contrast. (c) Image showing the peak positions identified. (d) An image of a single vortex created by averaging over all the vortex positions.

The criterion for selecting which peaks to manually exclude changes depending on whether one is interested in obtaining a good average image or investigating the structure of the vortex lattice. In the former, vortices where there is any unwanted contrast should be excluded. This was the criterion used in Figure 32 and it can be seen in panel (c) that some vortices which lie on a bend contour have been excluded in order to give a more accurate average. In the latter case, used in the correlation analysis in section 9.3, the best lattice will be obtained if as many of the positions are used as possible.

Figure 33 is offered as a visual comparison between simulated and experimental images and the two compare favourably. It should be noted that this comparison is rather unconstrained. The thickness of the sample was not measured and for more serious attempts at comparing theory and experiment, the convergent beam technique described in ref. 76 will probably prove to be the best method for determining the thickness. In the likely thickness range of 50-300 nm, simulations show that the contrast changes but visual appearance of the simulations hardly alters. The simulation presented here is for 200 nm.

The sample tilt can be measured to within 2° by assuming the vortex spacing is the same in all directions as discussed in section 9.2 and we found it to be 52 ± 2°. The magnification and defocus level were obtained by imaging a ruled carbon film directly after the experiment with the same lens settings (see section 8.2). Nevertheless, when the simulation was scaled to best fit the experimental image, the magnification of the simulation needed to be increased by 15%. This inconsistency is perhaps not too surprising as the magnification at this defocus is very sensitive to the defocus level and the C2 lens setting of the microscope. It is also unknown how accurate the calibration grid is and we intend to compare it with a catalase calibration sample.

It is also likely that the vortex is wider than allowed for in our simulations due to the spreading of the B-field at the sample surfaces described in section 6.3 as, in the likely thickness range, the specimen thickness is less than the penetration depth so the vortex is wider than it is long. This effect would also explain why the experimental contrast (defined as the difference between the maximum and minimum intensity divided by the sum of the maximum and minimum intensity) is 6%, considerably below the predicted values which range from 25% for a 50 nm specimen to 36% for a 300 nm specimen. Patti and Pozzi [60] give graphs showing the reduction in contrast when surface effects are accounted for and these show that it is entirely feasible that the reduction in contrast we observe is due to B-field spreading near the specimen surfaces. In the future we intend to make more accurate simulations using the Fourier space method described in ref 54 for comparison with specimens of known thickness.

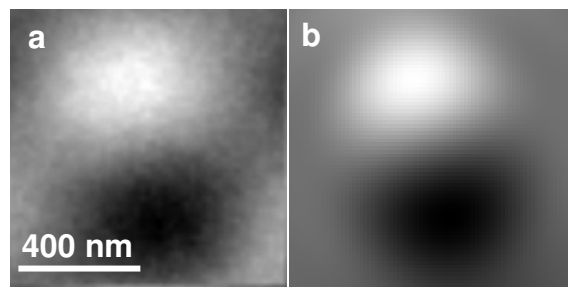

**Figure 33:** Visual comparison between experimental and simulated Fresnel Images. (a) Experimental image at 15 cm defocus obtained by averaging over many vortices as shown in Figure 32. (b) Simulation for a single vortex in BSCCO using $\xi$ = 3nm and $\Lambda$ = 300 nm for a 200 nm sample tilted to 52°. The maximum and minimum have been scaled to be the same in both images.

*9.2    Correction for Lattice Distortion*

One of the disadvantages of imaging vortices using electron microscopy is that the sample must be tilted and so the vortex lattice appears compressed in one direction. This compression can be observed in the power spectrum of the image: either as a distorted hexagon if the lattice has long range order or as an elliptical ring if the vortices adopt a 'polycrystalline' arrangement. An ellipse can be fitted to the power spectrum and the ratio of the minor to the major axes gives the cosine of the angle ($\alpha$) at which the specimen was tilted. In this case, the fit gave $\alpha = 52 \pm 2°$ where the error was judged by changing the ellipse from the maximum to its minimum acceptable fit. It should be noted that in the new stretched image, the black-white features are distorted but the lattice should be correct.

This procedure makes the assumption that the crystal is isotropic in the *a-b* plane. If the penetration depths are different in the *a* and *b* directions, the vortex lattice will be stretched in the *a* direction by $\Lambda_a/\Lambda_b$ since the circulating currents decay to zero over different lengths in *a* and *b* and so follow elliptical paths around the vortex core. In the case of BSCCO, the penetration depth is taken to be the same in *a* and *b* in all the literature we can find.

There are two further complications that should not affect this experiment but be considered in the general case. First, although the B-field and hence the repulsive force between vortices dies off over a length characterised by the penetration depth within the superconductor, the field outside the superconductor from the ends of the vortex decay more slowly. For widely spaced vortices, this field will be purely monopolar and the repulsion will be minimised if the vortex lattice is a regular hexagonal Wigner crystal. In the anisotropic case with different penetration depths in different directions, this would lead to an energy competition with the stray field repulsion favouring a regular hexagonal lattice and the field within the superconductor favouring a distorted hexagonal lattice.

Secondly, if the vortex is tilted as it passes through the superconductor, the *c* axis anisotropy must also be considered and the pattern of circulating currents determined using the 'effective mass tensor' as discussed in ref. 77. In our case, we expect the vortex to pass through conventional superconductors almost normal to the thin surface as discussed in section 4. In the case of BSCCO, the vortices are expected to form the crossing lattice shown in Figure 7(b) and the vortices we image are the pancake stacks with fields parallel to *c*. We expect the lattice of pancake stacks to be regular unless there is a Josephson vortex which causes the pancake vortices form 'vortex chains' [44]. Such chains have been observed by electron microscopy by Tonomura *et al.* [78] but we did not observe them here.

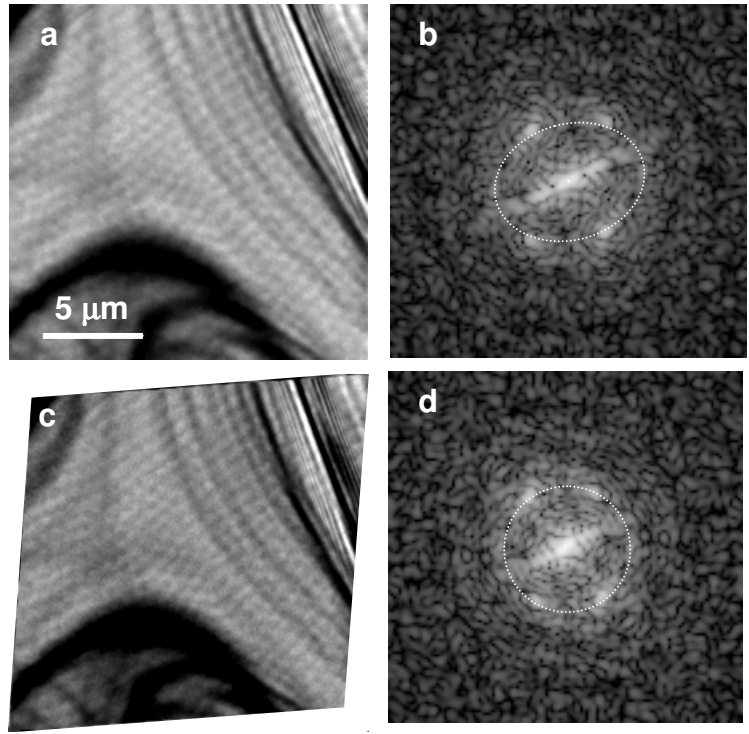

**Figure 34:** Correction for lattice distortion. (a) Original Fresnel image and (b) its power spectrum with an ellipse superimposed. (c) image compensated for the lattice distortion and (d) its power spectrum showing equal lattice parameters indicated by the circle superimposed.

*9.3  Correlation Analysis of Fluxon Lattice as a Function of Magnetic Field*

Flux vortices tend to form hexagonal arrays with varying degrees of disorder. Here we examine the methods for assessing how good the ordering is based on refs. 79, 80 and 81. The methods involve assigning an order parameter at the position of each vortex which describes how well ordered the surrounding, nearest-neighbour vortices are. The nearest neighbours for each vortex are assigned by Delaunay triangulation [82]. Having identified the nearest neighbours we wish to answer the following questions:

(1) If the average spacing of the vortices is $R_0$, by how much does the spacing of the vortices that surround a particular vortex deviate from $R_0$?

(2) The average angle between vortices is expected to be 60°. By how much do the angles of the vortices surrounding a particular vortex deviate from this ideal?

Both these questions have two components: how much does the average change and what is the spread about the average? This is illustrated in Figure 35.

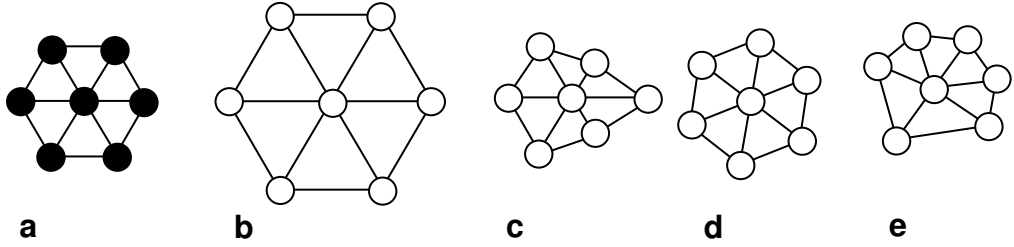

**Figure 35:** Lattice Distortions. (a) the perfect lattice. (b) a uniform lattice expansion. (c) the average lattice parameter is unchanged but there is disorder about the average. (d) a lattice rotation with respect to (a). (e) the lattice is in the same average orientation as (a) but there is a spread of angles about the ideal value.

There are many ways in which such deviations from the perfect lattice could be measured but it is conventional to measure these parameters in Fourier space [79] as we now describe.

Using polar coordinates $(r,\theta)$, we select a particular vortex and using this as an origin, the nearest neighbour vortices are given coordinates $(r_n,\theta_n)$. The density of the nearest-neighbour vortices is then written as

$$\rho(r,\theta) = \sum_{n=1}^{Z} \delta(r-r_n)\delta(\theta-\theta_n) \qquad (77)$$

where $Z$ is the number of nearest neighbour vortices surrounding the one in question so that

$$\int_{\theta=0}^{2\pi}\int_{r=0}^{\infty} \rho(r,\theta)r\,dr\,d\theta = Z \qquad (78)$$

We can now average radially and circumferentially to form two new functions which give the number of vortices per area between $r$ and $r + dr$ and the number per circumference length between $\theta$ and $\theta + d\theta$ respectively.

$$\rho(r) = \int_0^{2\pi} \rho(r,\theta)d\theta \qquad (79)$$

$$\rho(\theta) = \int_0^{\infty} \rho(r,\theta)dr \qquad (80)$$

Each of these functions can be broken down into Fourier components in $r$ and $\theta$.

$$P(k) = \int_0^{\infty} \rho(r)e^{2\pi ikr}dr \qquad (81)$$

$$P(\kappa) = \int_0^{\infty} \rho(\theta)e^{2\pi i\kappa\theta}d\theta \qquad (82)$$

Equation (82) can be immediately simplified because the density is single valued and thus repeats every time the angle goes from 0 to $2\pi$. The density can be written as

$$\rho(\theta) = \rho_1(\theta) * \sum_{l=-\infty}^{\infty} \delta(\theta - 2l\pi) \tag{83}$$

where $\rho_1 = \rho$ for $0 < \theta < 2\pi$ and is zero outside this range. Using the convolution theorem, the Fourier transform is then

$$P(\kappa) = P_1(\kappa) \sum_{m=-\infty}^{\infty} \delta(\kappa - m/2\pi) \tag{84}$$

$$P(\kappa) = \int_{-\infty}^{\infty} \rho_1(\theta) e^{2\pi i \kappa \theta} d\theta \left( \sum_{m=-\infty}^{\infty} \delta(\kappa - m/2\pi) \right) \tag{85}$$

The series of delta functions means the P is only non-zero when $\kappa = 2\pi/m$ so we can write

$$P(\kappa) = \sum_{m=-\infty}^{\infty} \left( \delta(\kappa - m/2\pi) \int_0^{2\pi} \rho(\theta) e^{im\theta} d\theta \right) \tag{86}$$

Thus $P(\kappa)$ is a series of delta functions and the strength of each one is given by the integral.

A more convenient alternative is to express the strength of each delta function with index $m$ as

$$P_m = \int_0^{2\pi} \rho(\theta) e^{im\theta} d\theta \tag{87}$$

The index $m$ tells you how many times the lattice repeats as $\theta$ goes from 0 to $2\pi$ and if we anticipate a hexagonal ordering, the strength of the $m = 6$ component gives an indication of how close the lattice comes to perfect 6-fold ordering.

We choose order parameters that give the strength of the Fourier component of each function which would be maximal if the lattice were perfect. In a perfect lattice, the spacing between a vortex and its neighbours would always be the same so if we call the average spacing $R_0$, we should examine the $k_0$ Fourier component where $k_0 = 1/R_0$. Similarly a perfect hexagonal lattice would always have $m = 6$ so we examine the strength of the $m = 6$ component. Taking this into account and normalising by the number of neighbouring vortices gives the spatial and angular order parameters:

$$O_{\text{spatial}} \equiv \frac{1}{Z} P(k_0) = \frac{1}{Z} \int_0^{2\pi} \rho(r) e^{2\pi i k_0 r} dr \tag{88}$$

$$O_{angular} \equiv \frac{1}{Z}P_6 = \frac{1}{Z}\int_0^{2\pi} \rho(\theta)e^{i6\theta}d\theta \qquad (89)$$

Since the densities are given by equation (77), the order parameters can be written more simply as

$$O_{spatial} = \frac{1}{Z}\sum_{n=1}^{Z} e^{2\pi i k_0 r_n} \qquad (90)$$

$$O_{angular} = \frac{1}{Z}\sum_{n=1}^{Z} e^{i6\theta_n} \qquad (91)$$

It can be seen by examination that the phase of the spatial order parameter indicates regions of the image where there has been a lattice expansion or contraction of the form shown in Figure 34(b) and the phase of the angular correlation function identifies regions where there has been a rotation of the lattice of the form shown in Figure 34(d). The moduli of the two functions express the deviation from equally spaced vortices (Figure 34(c)) and the deviation from vortices distributed with equal angles (Figure 34(e)) respectively. Figure 36 demonstrates how the phase of the angular order parameter can be used to identify different 'grains' in the vortex lattice.

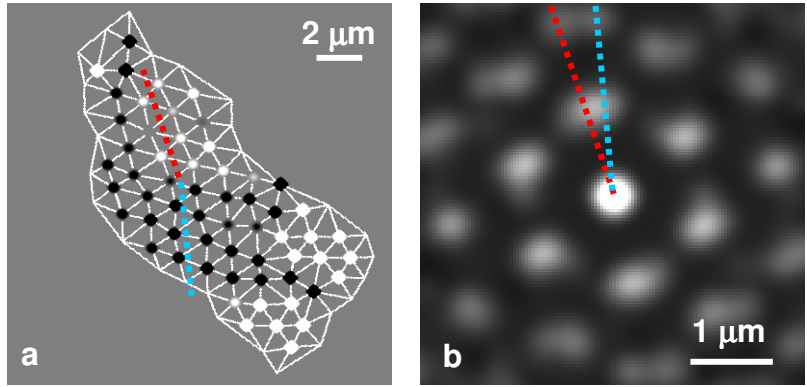

**Figure 36:** Identifying Grains using the Phase of the Angular Order Parameter. (a) The argument of the angular order parameter is shown as a greyscale value at each fluxon position and the fluxon positions are connected to their nearest neighbours using Delaunay triangulation. (Note that the greyscale values have been saturated to show the grains more clearly.) Two grains can be seen from the change from black to white and their orientation change is indicated by red and blue dotted lines. (b) The splitting of the spots in the correlation function of the angular order parameter also indicates that two grains are present.

In order to further investigate the crystallinity of the vortex lattice, it is also useful to identify the lengthscale over which the order parameter changes. This is done by making use of autocorrelation functions which give information on the extent to which one region of the vortex lattice resembles another.

If the position vector of each vortex is $\mathbf{R_i}$, the spatial distribution of the order parameter can be represented as a series of delta functions each weighted by the value of the order parameter at that point, $O_i$.

$$O(\mathbf{R}) = \sum_i O_i \delta(\mathbf{R} - \mathbf{R_i}) \tag{92}$$

We also broaden the point representing each fluxon position with a Gaussian broadening function with unit area, $f(\mathbf{R})$. If no broadening function is used, the lattice is a set of infinitesimal points and the correlation function is noisy. We choose the broadening function to be large enough to reduce the noise but not too large to obscure the salient details of the correlation function. The order parameter is then

$$O(\mathbf{R}) = f(\mathbf{R}) * \sum_i O_i \delta(\mathbf{R} - \mathbf{R_i}) \tag{93}$$

The distance over which the order parameter remains unchanged is assessed by calculating the autocorrelation function of the order parameter which is defined as

$$O(\mathbf{r}) \otimes O(\mathbf{r}) = \int O^*(\mathbf{R}) O(\mathbf{r} + \mathbf{R}) d^2\mathbf{R} \tag{94}$$

In this investigation, we normalise the autocorrelation as follows [83]:

$$G(\mathbf{r}) = \frac{\int_S O^*(\mathbf{R}) O(\mathbf{r} + \mathbf{R}) d^2\mathbf{R} - \left|\int_S O(\mathbf{R}) d^2\mathbf{R}\right|^2}{\int_S |O(\mathbf{R})|^2 d^2\mathbf{R} - \left|\int_S O(\mathbf{R}) d^2\mathbf{R}\right|^2} \tag{95}$$

It can be seen by inspection that this function yields a correlation which has a central peak with a value of 1 and an average of zero. When $G(\mathbf{r})$ has a value of 1, it indicates perfect correlation, a value of zero indicates no correlation and a value of -1 indicates perfect anticorrelation.

A measure of the length over which a function remains correlated is given by the correlation length, $\sigma$, which is obtained by first taking a circumferential average $G(\mathbf{r})$ to yield $G(r)$ and then comparing it with the correlation function that would be obtained if the lattice were perfect, $G_{\text{perfect}}(r)$. This is done by assuming that $G(r)$ can be related to $G_{\text{perfect}}(r)$ by multiplying by a decaying exponential function thus:

$$G(r) = G_{\text{perfect}}(r) \exp(-r/\sigma) \tag{96}$$

and using a least-squares fit to give the correlation length $\sigma$.

The imaginary part of the correlation tends to be considerably smaller than the real part and most investigators simply use the real parts of $G(r)$ for this fit. Here, the perfect lattice was obtained by using $G(r)$ to identify the average lattice parameter and orientation and then creating a perfect hexagonal lattice for an image of the same size with the same orientation and lattice parameter. From this perfect hexagonal lattice, the spatial and angular order parameters were obtained and then $G_{\text{perfect}}(r)$ was calculated. This procedure reduces the artefacts from the finite size of the image since both the image and the perfect lattice are subject to similar boundary conditions.

Figure 37 shows how the vortex lattice at 11 K in BSCCO changes as the applied B-field is reduced. Note that the values of the applied B-field are rather nominal as discussed in section 8.2 and it is likely that the position of zero field is inaccurate although changes in field should be approximately correct. At any rate, the lattice becomes sparser as the field is reduced as the fluxons move towards the specimen edge and disappear. The order in the lattice is created from the mutual repulsion between the vortices and as they get further apart, the repulsive force diminishes and the lattice becomes more disordered.

On a large scale, the correlation length should decrease monotonically with field but on the mesoscopic scale of the observations made here, there is considerable variability as illustrated in Figure 38 which uses the same data set as Figure 37. Figure 38(a) shows the trapped flux density – the number of vortices per unit area multiplied by the flux per vortex ($\Phi_0$) – as a function of B-field. In an equilibrium situation with no pinning forces, this should give a straight line with unit gradient but here the graph shows considerable variability and the best fit straight line has a gradient of $0.45 \pm 0.05$.

The reasons for this departure from the ideal are twofold: first pinning forces trap the vortices so that the number in the sample can be out of equilibrium with the applied B-field; secondly, it takes time for the vortices to rearrange themselves into a stable configuration after each field increment. Videos taken during the experiments indicated that it took between 10 seconds and 1 minute for the vortex lattice to stabilise. A good example of this is shown in Figure 37(d) and (e) where the arrangement changes considerably in the space of 20 seconds although the applied field remained the same. The same two points can be seen in Figure 38 at an applied field of -1.8 G. Panel (a) shows that the vortex lattice in this region of specimen becomes denser after the rearrangement and (b) and (c) show an increase in both the spatial and angular correlation lengths.

Another feature of the series we ascribe to the phenomenon of 'matching fields' [84]. In experiments concerned with the arrangements of vortices in periodic arrangements of defects, it is found that the vortex lattice is most ordered when the field is such that the periodicity of the vortices is a simple multiple of the periodicity of the defects. This concept presumably applies to irregular arrangements of defects which will inevitably be present in this sample: there will be certain fields at which the vortex lattice can fit among the defects and others at which the lattice is strained.

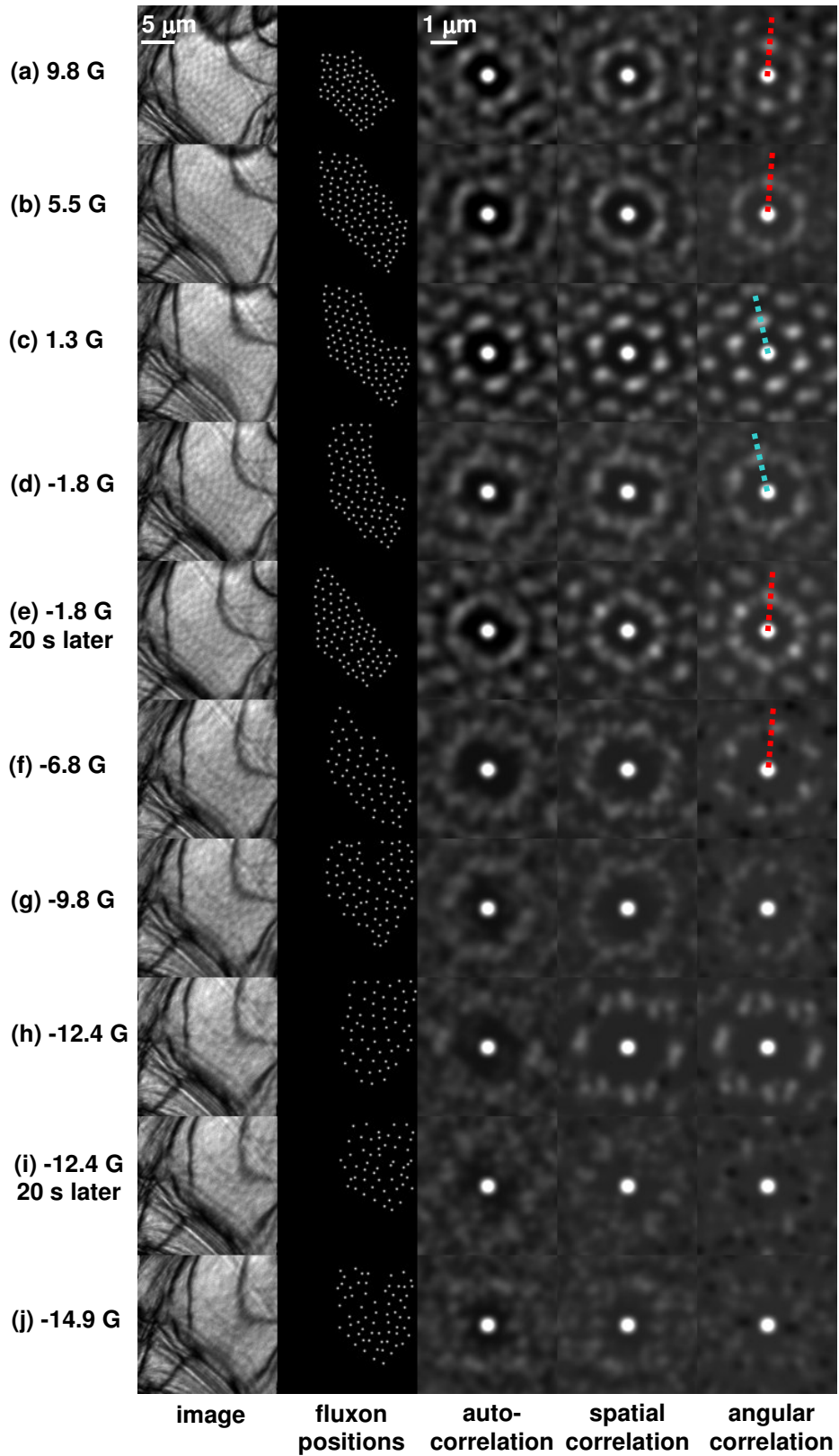

**Figure 37:** Correlation functions as a function of applied B-field at 11 K. The standard autocorrelation function in the third column is obtained by weighting each vortex position equally prior to calculating the autocorrelation. The calculation of the spatial and angular correlation functions is described in the text.

In this series, the lattice at 1.3 G is better ordered than at other fields. The orientation of the correlation function is interesting in this respect too. It can be seen that the lattice's principal axis starts at an orientation indicated by the red line in Figure 37. At a field of 1.3 G, the principal axis rotates by $15 \pm 1°$ and now points in the direction indicated by the blue line. This orientation is retained until the field reaches -1.8 G at which point (20 seconds after the field increment) it returns to its *original* orientation. We attribute this memory of the original orientation to the arrangement of pinning sites in the specimen.

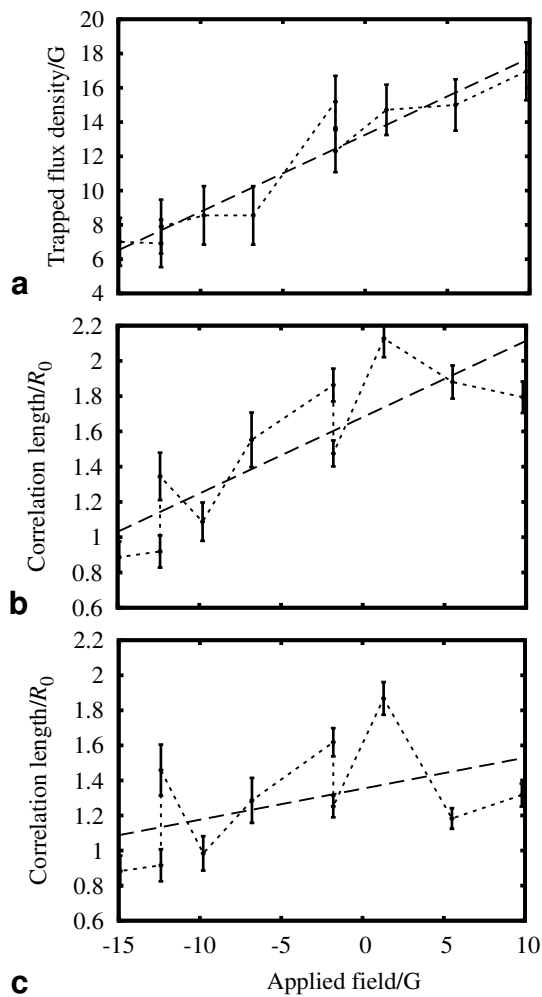

**Figure 38:** (a) Trapped flux as a function of applied B-field. (b) Spatial and (c) angular correlation lengths given as a multiple of the average vortex spacing, $R_0$ as a function of applied B-field. The error bars reflect the variation in the vortex spacing about the average value. The dashed lines are a guide to the eye showing the overall trends and the dotted lines show the order in which the measurements were taken beginning at the highest B-field.

As shown in Figure 38, the angular correlation length is less than the spatial correlation length which is the opposite to what is normally observed [79]. This means that the spacing between adjacent vortices remains similar over a longer lengthscale than the angle of the hexagons surrounding each vortex. This could be an artefact of the small numbers of vortices in the images but it is more likely that many grains are present in the lattice and whereas the average spacing remains similar, the orientation of the hexagonal order is variable.

## 10.  Conclusions and Future Work

Transmission electron microscopy provides a useful addition to other techniques for imaging flux vortices in superconductors. Unlike almost all other imaging techniques, it has the ability to probe the internal structure of vortices as it is sensitive to B-fields throughout the thickness of the sample not just at the surface. In addition, its spatial and temporal resolution and B-field sensitivity are as good as or superior to other techniques as summarised in Table I.

We have demonstrated that imaging vortices does not require a dedicated microscope and we have taken Fresnel images of vortices with a commercially available microscope equipped with a field emission gun and Lorentz lens. The main difficulty with the technique is the preparation of samples with large (>30 µm) electron transparent regions. Preparing flat samples is also very important otherwise diffraction contrast from sample bending tends to dominate Fresnel images, obscuring the vortices. Contrast from defects also tends to obscure the vortices and at present the only samples which have been studied by any researcher to date are single crystals or polycrystals with very large grains (~100 µm).

Simulations indicate that whilst Fresnel imaging is probably the simplest method for imaging vortices, the images do not contain much information on the internal structure of individual vortices and that even the image from an vortex of infinitesimal radius is very similar to that from a vortex with a realistic width at the defoci needed to give sufficient contrast to observe vortices.

On the other hand, simulations show that the diffraction pattern is very sensitive to the internal structure of the vortices and in the future we intend to take diffraction patterns for comparison with simulations. The main problem appears to be dealing with the dynamic range of the diffraction patterns (which simulations indicate is ~$10^6$) and thus imaging plates may be the best recording medium.

Simulations also show that imaging vortices with the use of a phase plate placed in the back focal plane produce vortex images with a higher contrast than Fresnel images and a much better spatial resolution. We have simulated the effect of a Hilbert phase plate and find that a π shifting phase plate is optimal although this can vary by as much as 1 radian without undue loss of contrast. For BSCCO it should be placed within 0.5 µrad of the most intense part of the diffraction pattern. This appears to be a very promising method for imaging vortices and given the higher contrast of the vortices, it may be possible to image vortices in more defective, commercially relevant samples such as thin films grown on substrates.

We have successfully taken Fresnel images from superconducting BSCCO and find that the images of individual vortices are in good agreement with simulations. We have demonstrated that once the image of a single vortex has been identified, this can be cross-correlated with the image to automatically pick out the other vortices in the image. This has the double benefit that the positions of the vortices can be identified and that the images can be averaged over every vortex to produce an image with a better signal to noise ratio.

Once the vortex positions have been identified, the spatial and angular correlation functions can be calculated and these provide a wealth of information on changes in the lattice structure as the external conditions applied to the vortices are altered. The ability to perform this type of correlation analysis is not restricted to transmission electron microscopy and indeed electron microscopy is not ideally suited because the field of view is limited to ~30 µm, restricting the number of vortices which can be investigated. There is also a geometric distortion introduced by the sample tilt which must be corrected.

For the analysis of static lattices, the Bitter technique is best suited as it has an almost unlimited field of view and the great majority of correlation analyses have been done using this technique (see e.g ref. 72). For dynamic studies, it would seem from Table I that magneto-optical imaging is best as it gives a wide field of view (~mm) and has good temporal resolution (~0.1 s). Unexpectedly, no similar correlation analyses have been undertaken to our knowledge although the ability to image single vortices using magneto-optical imaging was demonstrated in 2001 [31].

It is also possible to obtain correlation functions from vortex lattices without starting from an image. It is well known that the pair distribution function of a lattice, which is essentially the autocorrelation function with the central peak removed, can be obtained from a Fourier inversion of a diffraction pattern as described in ref. 81. A variant on this technique is described in ref. 85 where a Monte-Carlo technique is used to generate model vortex lattices that would fit a given neutron diffraction pattern. This technique does not produce a unique vortex lattice but rather a series of lattices consistent with the diffraction pattern but the correlation functions from each of these lattices are the same. The diffraction patterns are taken from millimetre sized crystals and this is effectively the field of view. The disadvantage of the technique is that the weak interaction of neutrons with vortices mean that long acquisition times ranging from tens of minutes to several hours are needed to collect the data so that the temporal resolution is poor. Its great advantage is its large field of view but this is also a disadvantage as data cannot also be taken from small regions of the crystal.

Despite the restricted field of view, our results and those of refs. 73 and 74 demonstrate that a correlation analysis can be profitably performed using electron micrographs although the real benefit in using transmission electron microscopy lies in its ability to perform several different analyses at the same time. We have shown that the same images can be used both for analysing the structure of the vortex lattice and also the structure of individual vortices within the lattice and that image sequences can be acquired of dynamic processes with a time resolution of video rate to a few seconds.

We illustrate the simultaneous application of these techniques by imagining an experiment where a video of flux creep is analysed. A correlation analysis could be performed on each frame giving a map showing grains with different lattice orientations and regions of lattice expansion and contraction as well as identifying lattice defects and how all of these change with time. The structure of individual vortices could be investigated from the same data set and if the structure of individual vortices varied because, for example, they were pinned as in ref. 61, a variation on the motif averaging procedure would allow different types of vortex to be automatically identified. An analysis of the flow of vortices would allow the pinning sites to be identified as in ref. 73 and their effect on the vortex flow ascertained. One could then use conventional electron microscopy to investigate the nature of the defects contributing to each pinning site. Investigations of this variety would illuminate the types of pinning site which impede the vortex flow best, potentially giving an insight into how superconducting devices can be designed to minimise power loss and noise. As well as this, this type of analysis will give a valuable insight into such phenomena as the vortex ratchet effect where pinning sites are arranged so that vortices flow preferentially in one direction [86] and a real time view of the vortex lattice melting transition thought to occur in BSCCO where, under certain conditions, the pinning forces on vortices seem to disappear [87].

## 11. Acknowledgements

We should like to thank Prof. J.R. Cooper, Dr. T. Benseman and C. L. Zentile of the Quantum Matter Group in the Cavendish Laboratory, University of Cambridge who provided the BSCCO samples used in this investigation. They also provided advice on superconductivity as did as C. Bowell, formerly a member of the Condensed Matter Physics Group at the University of Birmingham and now a member our group. We also thank Dr R.E. Dunin-Borkowski for advice and providing the B-field calibration used here as well as Prof G. Pozzi and Dr M. Beleggia for stimulating discussions. This investigation was funded by the Royal Society, the EPSRC and Homerton College, University of Cambridge. PAM acknowledges financial support from the European Union under the Framework 6 program for an Integrated Infrastructure Initiative, Ref.: 026019 ESTEEM.

## 12. Appendix: Phase Shift from a Tilted Vortex of Infinitesimal Radius

In section 6.1 we derived the phase shift for a vortex of infinitesimal radius lying normal to the electron beam. Using the same geometry, we now extend the derivation to give the phase shift from a vortex in a specimen tilted as shown in Figure A1.

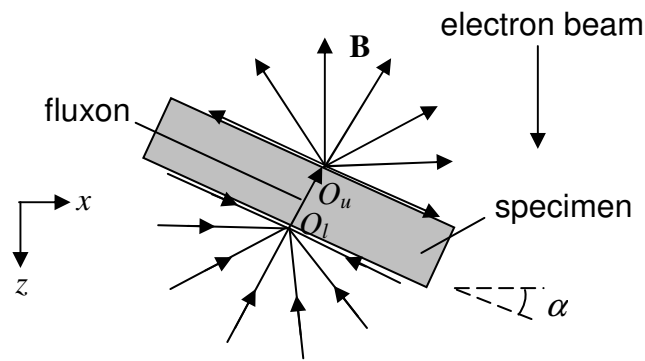

**Figure A1:** Schematic B-field lines from an infinitesimally narrow flux vortex in a specimen tilted at an angle $\alpha$ with respect to the horizontal $xy$ plane.

Migliori *et al.* [49] treat the case of a tilted fluxon by first considering the phase shift from the upper monopole in a semi-infinite specimen and we follow their derivation. The phase shift from the upper monopole at $O_u$ (see Figure A1) is still equal to negative one half of the solid angle subtended by the plane defined by the trajectories $\Gamma_0$ and $\Gamma$ and the specimen plane as derived in section 6.1: the problem is finding the solid angle. The solid angle can be visualised as the area of the spherical triangle PSP$_0$ outlined in dots in Figure A2 which is part of a sphere with unit radius centred at $O_u$.

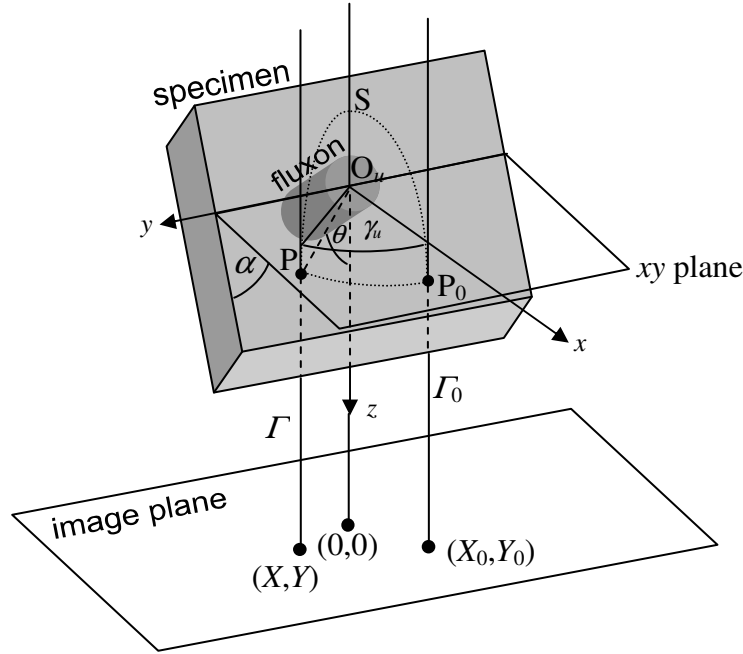

**Figure A2:** Geometry for calculating the phase shift from the upper monopole centred at $O_u$. The reference trajectory $\Gamma_0$ passes through the *xy* plane and meets the specimen at point P$_0$. It then intersects the image at point $(X_0,Y_0)$ where the phase shift is defined to be zero. The relative phase shift at another point in the image $(X,Y)$ is evaluated along the trajectory $\Gamma$ which passes through a point P on the specimen. As discussed in the text, the problem of finding the phase shift is equivalent to finding the area of the spherical triangle PSP$_0$ which is part of a sphere of unit radius centred on $O_u$.

Referring to Figure A2, the equation of the plane of the specimen is

$$z = x \tan \alpha$$

The reference trajectory $\Gamma_0$ which defines the point of zero phase shift (see section 6.1) is taken to be $\gamma_u = 0$.

Thus the point P is where the second trajectory $\Gamma$ hits the specimen plane has position vector (normalised to have length 1)

$$\begin{pmatrix} x \\ y \\ z \end{pmatrix} = \frac{1}{\sqrt{1+\cos^2 \gamma_u \tan^2 \alpha}} \begin{pmatrix} \cos \gamma_u \\ \sin \gamma_u \\ \cos \gamma_u \tan \alpha \end{pmatrix}.$$

We now find the polar angle of point P, $\theta$, measured from the $z$ axis by taking the dot product of the position vector of P with the $z$ axis, (0,0,1).

$$\cos\theta = \begin{pmatrix}0\\0\\1\end{pmatrix}\cdot\begin{pmatrix}\cos\gamma_u\\ \sin\gamma_u\\ \cos\gamma_u\tan\alpha\end{pmatrix}\frac{1}{\sqrt{1+\cos^2\gamma_u\tan^2\alpha}} = \frac{\cos\gamma_u\tan\alpha}{\sqrt{1+\cos^2\gamma_u\tan^2\alpha}}$$

The element of solid angle is then given by

$$d\Omega = \sin\theta\, d\theta\, d\gamma_u = -d(\cos\theta)d\gamma_u$$

So the solid angle as a function of $\gamma_u$ and $\theta$ is

$$\Omega = \int_{\gamma'=0}^{\gamma_u}\int_{\theta'=\theta}^{\pi} -d(\cos\theta')d\gamma'$$

$$= \int_0^{\gamma_u}(1+\cos\theta)d\gamma'$$

$$= \int_0^{\gamma_u}\left(1+\frac{\cos\gamma'\tan\alpha}{\sqrt{1+\cos^2\gamma'\tan^2\alpha}}\right)d\gamma'$$

$$= \int_0^{\gamma_u}\left(1+\frac{\cos\gamma'\tan\alpha}{\sqrt{1+(1-\sin^2\gamma')\tan^2\alpha}}\right)d\gamma'$$

$$= \int_0^{\gamma_u}\left(1+\frac{\cos\gamma'\tan\alpha}{\sqrt{1+\tan^2\alpha-\sin^2\gamma'\tan^2\alpha}}\right)d\gamma'$$

If we let $Z = \sin\gamma'\tan\alpha \Rightarrow dZ = \cos\gamma'\tan\alpha\, d\gamma'$, the integral becomes

$$\Omega = \gamma_u + \int_0^{\sin\gamma_u\tan\alpha}\frac{dZ}{\sqrt{(1+\tan^2\alpha)-Z^2}}$$

$$= \gamma_u + \left[\sin^{-1}\left(\frac{Z}{\sqrt{1+\tan^2\alpha}}\right)\right]_0^{\sin\gamma_u\tan\alpha}$$

$$= \gamma_u + \sin^{-1}\left(\frac{\sin\gamma_u\tan\alpha}{\sqrt{1+\tan^2\alpha}}\right)$$

$$\Omega = \gamma_u + \sin^{-1}(\sin\gamma_u\sin\alpha)$$

So the phase shift for the upper monopole is given by

$$\phi_u = -\frac{1}{2}\left(\gamma_u + \sin^{-1}\left(\sin\gamma_u \sin\alpha\right)\right)$$

As in section 6.1, the overall phase shift from a specimen can be written as a sum of two terms. The first term is the phase shift from a semi-infinite specimen that we have just derived and the second term is similar in form giving the opposite phase shift in the vacuum beneath the lower monopole centred at $O_l$ and cancelling the phase jump of $\pi$ in this region produced by the first term.

$$\phi = -\frac{1}{2}\left(\gamma_u + \sin^{-1}\left(\sin\gamma_u \sin\alpha\right)\right) + \frac{1}{2}\left(\gamma_l - \sin^{-1}\left(\sin\gamma_l \sin\alpha\right)\right)$$

## 13. References

1. W. Meissner and R. Ochsenfeld, Naturwissenschaften 21 (1933) 787.
2. J. Bardeen, L. N. Cooper, J. R. Schrieffer, Phys. Rev. 108 (1957) 1175.
3. J.F. Annett, Superconductivity, Superfluids and Condensates, O.U.P., Oxford, 2004.
4. N.W. Ashcroft and N.D. Mermin, Solid State Physics, Saunders College Publishing, USA, 1976.
5. F. London and H. London, Proc. Roy. Soc, Lond. A149 (1935) 71.
6. C.G. Kuper, An Introduction to the Theory of Superconductivity, Clarendon Press, Oxford, 1968.
7. G.W.C. Kaye, T.H. Laby, Tables of Physical and Chemical Constants, Longman, London, 1973.
8. J. Evetts (ed.), The Concise Encyclopedia of Magnetic and Superconducting Materials, Pergamon, Oxford, 1992.
9. A.A. Abrikosov, Soviet Physics JETP 5 6 (1957) 1174.
10. P.B. Hirsch, A. Howie, J.P. Jakubovics, Proceedings of the International Conference on Electron Diffraction: Crystal Defects, 1 B-5, Pergamon, Oxford, 1965.
11. H. Yoshioka, J. Phys. Soc. Jap. 21 (1966) 948.
12. D. Wohlleben, J. Appl. Phys. 38 8 (1967) 3341.
13. C. Colliex, B. Jouffrey, M. Kleman, Acta Cryst. A24 (1968) 692.
14. C. Capiluppi, G. Pozzi, U. Valdrè, Phil. Mag. 26 (1972) 865.
15. T. Matsuda, S. Hasegawa, M. Igarashi, T. Kobayashi, M. Naito, H. Kajiyama, J. Endo, N. Osakabe, A. Tonomura, R. Aoki, Phys. Rev. Lett. 62 21 (1989) 2519.
16. M. Tinkham, Phys. Rev. 129 6 (1963) 2413.
17. K. Harada, T. Matsuda, J. Bonevich, M. Igarashi, S. Kondo, G. Pozzi, U. Kawabe, A. Tonomura, Nature 360 (1992) 51.
18. T. Kawasaki, T. Matsuda, J. Endo, A. Tonomura, Jpn. J. Appl. Phys. 29 3 (1990) 508.
19. K. Harada, H. Kasai, T. Matsuda, M. Yamashaki, J.E. Bonevich, A. Tonomura, Jpn. J. Appl. Phys. 33 1 No. 5a (1994) 2534.
20. T. Yoshida, J. Endo, K. Harada, H. Kasai, T. Matsuda, O. Kamimura, A. Tonomura, M. Beleggia, R. Patti, G. Pozzi, J. Appl. Phys. 85 8 (1999) 4096.

21. S. Horiuchi, M. Cantoni, M. Uchida, T. Tsuruta, Y. Matsui, Appl. Phys. Lett. 73 9 (1998) 1293.
22. S. Horiuchi, M. Cantoni, M. Uchida, T. Tsuruta, Y. Matsui, Micron 30 (1999) 485.
23. T. Kawasaki, I. Matsui, T. Yoshida, T. Katsuta, S. Hayashi, T. Onai, T. Furutsu, K. Myochin, M. Numata, H. Mogaki, M. Gorai, T. Akashi, O. Kamimura, T. Matsuda, N. Osakabe, A. Tonomura, K. Kitazawa, J. Elec. Micros. 49 6 (2000) 711.
24. A. Tonomura, J. Elec. Micros. 52 1 (2003) 11.
25. A. Tonomura, H. Kasai, O. Kamimura, T. Matsuda, K. Harada, T. Yoshida, T. Akashi, J. Shimoyama, K. Kishio, T. Hanaguri, K. Kitazawa, T. Matsui, S. Tajima, N. Koshizuka, P.L. Gammel, D. Bishop, M. Sasase, S. Okayasu, Phys. Rev. Lett. 88 23 (2002) 237001.
26. P.E. Goa, H. Hauglin, M. Baziljevich, E. Il'yashenko, P.L. Gammel, T.H. Johansen, Supercond. Sci. Technol. 14 (2001) 729.
27. C.D. Dewhurst, R. Cubitt, Physica B 385-386 (2006) 176.
28. E.N. Kaufmann (ed.), Characterisation of Materials, Wiley, Hoboken, 2003.
29. A. Oral, S.J. Bending, M. Henini, Appl. Phys. Lett. 69 9 (1996) 1324.
30. S. Okayasu, T. Hishio, Y. Hata, J. Suzuki, I. Kakeya, K. Kadowaki, V.V. Moshchalkov, IEEE Transactions on Applied Superconductivity 15 2 (2005) 696.
31. P.E. Goa, H. Hauglin, Å.A.F. Olsen, M. Baziljevich, T.H. Johansen, Review of Scientific Instruments 74 (2003) 141.
32. R. Proksch, G. D. Skidmore, E. D. Dahlberg, S. Foss, C. Merton, B. Walsh, Appl. Phys. Lett. 69 17 (1996) 2599.
33. M. Rohmer, C. Wiemann, M. Munzinger, L. Guo, M. Aeschlimann, M. Bauer, Appl. Phys. A 82 (2006) 87.
34. R. Curtis, T. Mitsui, E. Ganz, Rev. Sci. Instrum. 68 7 (1997) 2790.
35. H. Nagaoka, Phil. Mag. 6 (1903) 19.
36. A.G. Webster, Bulletin Am. Math. Soc. July (1907) 1.
37. R. Lundin, Proc. IEEE, 73 9 (1985) 1428.
38. A. Yu. Martynovich, Zh. Eksp. Teor. Fiz. 105 (1994) 912.
39. E.H. Brandt, Phys. Rev. B 49 9 (1993) 6699.
40. M. Beleggia, G. Pozzi, J. Masuko, N. Osakabe, K. Harada, T. Yoshida, O. Kamimura, H. Kasai, T. Matsuda, A Tonomura, Phys. Rev. B 66 (2002) 174518.
41. E. H. Brandt, Phys. Rev. B 48 9 (1993) 6699.
42. G. Blatter, M.V. Feigel'man, V.B. Geshkenbein, A.I. Larkin, V.M. Vinokur, Rev. Mod. Phys. 66 4 (2004) 1125.
43. M. Beleggia, G. Pozzi, A. Tonomura, H. Kasai, T. Matsuda, K. Harada, T. Akashi, T. Matsui, S. Tajima, Phys. Rev. B 70 (2004) 184518.
44. A.E. Koshelev, Phys. Rev. Lett. 83 1 (1999) 187.
45. M. Beleggia, Phys. Rev. B 69 (2004) 014518.
46. A. Tonomura, J. Elec. Micros. 52 1 (2003) 11.
47. R.F. Egerton, Electron Energy-Loss Spectroscopy in the Electron Microscope, 2nd edtn., Plenum Press, New York, 1996.
48. A. Migliori, G. Pozzi, Ultramicroscopy 41 (1992) 169.
49. A. Migliori, G. Pozzi, A. Tonomura, Ultramicroscopy 49 (1993) 87.

50. S. Fanesi, G. Pozzi, J.E. Bonevich, O. Kamimura, H. Kasai, K. Harada, T. Matsuda, A. Tonomura, Phys. Rev. B 59 2 (1999) 1426.
51. J. R. Clem, J. Low Temp. Phys. 18 (1975) 427.
52. W. H. Press, S. A. Teukolsky, W. T. Vetterling, B. P. Flannery, Numerical Recipies in Fortran 77 second edtn., C.U.P., Cambridge, 2003.
53. J. Bonevich, D. Capacci, K. Harada, H. Kasai, T. Matsuda, R. Patti, G. Pozzi, A. Tonomura, Phys. Rev. B, 57 22 (1998) 1200.
54. M. Beleggia and G. Pozzi, Phys. Rev. B 63 (2001) 054507.
55. J.E. Bonevich, K. Harada, H. Kasai, T. Matsuda, T. Yoshida, G. Pozzi, Phys. Rev. B 49 10 (1994) 6800.
56. M. Beleggia and G. Pozzi, Ultramicroscopy 84 (2000) 171.
57. T. Yoshida, J. Endo, H. Kasai, K. Harada, N. Osakabe, A. Tonomura, G. Pozzi, J. Appl. Phys. 85 2 (1999) 1228.
58. K. Harada, H. Kasai, T. Matsuda, M. Yamasaki, J.E. Bonevich, A. Tonomura, Jpn. J. Appl. Phys. 33 (1994) 2534.
59. P.B. Hirsch, A. Howie, R. Nicholson, D.W. Pashley, M.J. Whelan, Electron Microscopy of Thin Crystals, 2nd edtn., Krieger, Malabar, Florida, 1977.
60. R. Patti, G. Pozzi, Ultramicroscopy 77 (1999) 163.
61. M. Beleggia, G. Pozzi, J. Masuko, N. Osakabe, K. Harada, T. Yoshida, O. Kamimura, H. Kasai, T. Matsuda, A. Tonomura, Phys. Rev. B 66 (2002) 174518.
62. A.B. Johnston and J.N. Chapman, J. Micros. 179 2 (1995) 119.
63. R. Danev and K. Nagayama, Ultramicroscopy 88 (2001) 243.
64. M. Beleggia, Ultramicrscopy 108 (2008) 953.
65. A.J.C. Wilson, International Tables for Crystallography, vol. C, Kulwer Academic Publishers, London 1992.
66. P.A. Midgley, Micron 32 2 (2001) 167.
67. H. Lichte, K.-H. Herrmann, F. Lenz, Optik 77 3 (1987) 135.
68. D. van Dyck, J. Elec. Micros. 48 1 (1999) 33.
69. T.M. Benseman, J.R. Cooper, G. Balakrishnan, Physica C 468 (2008) 81.
70. A. Tonomura, H. Kasai, O. Kamimura, T. Matsuda, K. Harada, T. Yoshida, T. Akashi, J. Shimoyama, K. Kishio, T. Hanaguri, K. Kitazawa, T. Matsui, S. Tajima, N. Koshizuka, P.L. Gammel, D. Bishop, M. Sasase, S. Okayasu, Phys. Rev. Lett., 88 23 (2002) 237001.
71. J.W. Lau, M.A. Schofield, Y. Zhu, Ultramicroscopy 107 (2007) 396.
72. C.A. Murray, P.L. Gammel, D.J. Bishop, D.B. Mitzi, A. Kapitulnik, Phys. Rev. Lett. 64 19 (1990) 2312.
73. C.-H. Sow, K. Harada, A. Tonomura, G. Crabtree, D.G. Grier, Phys. Rev. Lett, 80 12 (1998) 2693.
74. M. Mungan, G.-H. Sow, S.N. Coppersmith, D.G. Grier, Phys. Rev. B 58 21 (1998) 14588.
75. W. O. Saxton, J. Struc. Biol. 116 (1996) 230.
76. Williams D.B. and Carter C.B., Transmission Electron Microscopy, chapter 21, Plenum Press, New York, 1996.
77. S.L. Thiemann, Z. Radović, V.G. Kogan, Phys. Rev. B 39 16 (1989) 11406.
78. A. Tonomura, H. Kasai, O. Kamimura, T. Matsuda, K. Harada, T. Yoshida, T. Akashi, J. Shimoyama, K. Kishio, T. Hanaguri, K. Kitazawa, T. Matsui, S.

Tajima, N. Koshizuka, P.L. Gammel, D. Bishop, M. Sasase, S. Okayasu, Phys. Rev. Lett. 88 23 (2002) 237001.
79. D.R. Nelson, B.I. Halperin, Phys. Rev. B 19 5 (1979) 2457.
80. D.R. Nelson, M. Rubinstein, F. Spaepen, Phil. Mag. A46 1 (1982) 105.
81. K. Binder, W. Kob, Glassy Materials and Disordered Solids, chapter 2, World Scientific Publishing Company, Singapore, 2005.
82. W.H. Press, S.A. Teukolsky, W.T. Vetterling, B.P. Flannery, Numerical Recipes, The Art of Scientific Computing, 3rd edtn., Chapter 21, C.U.P., Cambridge, 2007.
83. W.O. Saxton, J. Micros. 190 1/2 (1998) 52.
84. A.N. Lykov, Sol. Stat. Comm. 86 8 (1993) 531.
85. M. Laver, E.M. Forgan, A.B. Abrahamsen, C. Bowell, Th. Geue, R. Cubitt, Phys. Rev. Lett. 100 (2008) 107001.
86. Y. Togawa, K. Harada, T. Asashi, H. Kasai, T. Matsuda, A. Maeda, A. Tonomura, Physica C 426–431 (2005) 141.
87. A. Soibel, E. Zeldiv, M. Rappaport, Y. Myasoedov, T. Tamegai, S. Ooi, M. Konczykowski, V.B. Geshkenbein, Nature 406 (2000) 282.